\documentclass[aps,prd,twocolumn,10pt,nofootinbib,superscriptaddress,amsmath,amssymb,longbibliography,floatfix]{revtex4-2}
\usepackage{graphicx,booktabs,amsmath,amssymb,mathtools,bm,array}
\usepackage[dvipsnames]{xcolor}
\usepackage{silence}
\usepackage{hyperref}
\usepackage{orcidlink}
\makeatletter
\newenvironment{plainwidetext}{%
\par\ignorespaces\onecolumngrid\vspace{2pt}%
}{%
\par\vspace{3pt}\twocolumngrid\global\@ignoretrue\@endpetrue%
}
\makeatother
\newcommand{\inlinewidefigure}[4]{%
\begin{plainwidetext}
\begin{center}
\vspace{-2pt}%
\includegraphics[width=#1\textwidth]{#2}\par
\vspace{-4pt}%
\refstepcounter{figure}\label{#4}%
\begin{minipage}{0.94\textwidth}
\footnotesize\textbf{FIG. \thefigure.} #3
\end{minipage}%
\vspace{-4pt}%
\end{center}
\end{plainwidetext}%
}
\definecolor{LinkBlue}{RGB}{0,55,160}
\definecolor{CiteMagenta}{RGB}{140,0,110}
\definecolor{URLTeal}{RGB}{0,90,125}
\hypersetup{colorlinks=true,linkcolor=LinkBlue,citecolor=CiteMagenta,urlcolor=URLTeal,filecolor=LinkBlue,anchorcolor=LinkBlue,menucolor=LinkBlue,breaklinks=true,pdfencoding=auto,psdextra}
\newcommand{\mpl}{M_{\rm Pl}}
\newcommand{\dd}{\mathrm{d}}
\newcommand{\ee}{\mathrm{e}}
\newcommand{\Order}{\mathcal{O}}
\newcommand{\cV}{\mathcal{V}}
\newcommand{\cs}{\rm cs}

\newcommand{\tw}{\rm tw}

\newcommand{\orcidicon}[1]{\orcidlink{#1}}
\begin{document}

\title{Penumbral Inflation from Calabi-Yau Boundaries}

\author{Pirzada \orcidicon{0009-0002-2274-9218}}%
\email{pirzada@itp.ac.cn}%
\affiliation{CAS Key Laboratory of Theoretical Physics, Institute of Theoretical Physics, Chinese Academy of Sciences, Beijing 100190, China}
\affiliation{School of Physical Sciences, University of Chinese Academy of Sciences, No. 19A Yuquan Road, Beijing 100049, China}

\author{Tianjun Li}
\email{tli@itp.ac.cn}
\affiliation{School of Physics, Henan Normal University, Xinxiang 453007, P. R. China}

\begin{abstract}
Mapping Calabi-Yau geometry to early Universe cosmology remains a primary goal in string phenomenology. We present an inflationary mechanism where flux stationarity shifts the field dynamics, driving inflation along the volume-controlling saxion direction near a Hodge boundary. In this region, the internal geometry strictly dictates the scalar potential. This setup stabilizes the field against heavy moduli corrections without consuming extra flux tadpoles. As a result, the properties of the geometric boundary translate directly into primordial tensor perturbations, creating a clear physical link from string compactifications to measurable data. The resulting geometric predictions match current cosmic microwave background limits and provide specific, testable targets for upcoming LiteBIRD and CMB-S4 observations. 
\end{abstract}
\maketitle

\section{Introduction}

Inflation explains accelerated expansion and relates primordial perturbations to quantum fluctuations~\cite{Guth:1981,Linde:1982}. The nearly scale invariant scalar spectrum follows from the quantum dynamics of the inflationary background~\cite{Starobinsky:1980,Mukhanov:1981}. Current CMB measurements constrain the tensor to scalar ratio and the inflationary energy scale~\cite{Planck:2020,BICEPKeck:2021,Tristram:2021}. LiteBIRD and CMB-S4 target the range near $r\sim10^{-3}$~\cite{LiteBIRD:2022,CMBS4:2022}. In a string boundary construction, a signal in this range also probes the canonical slope generated by a logarithmic metric pole. Limiting Hodge structure determines the pole residue~\cite{Schmid:1973,CattaniKaplanSchmid:1986,GrimmPaltiValenzuela:2018}. Boundary geometry does not by itself identify the field that traverses the pole. A canonical plateau therefore omits the flux orbit and the massive stationary directions that determine the cosmological tangent. Predictivity requires compactification data to determine both the metric and the trajectory.

Type IIB compactifications support inflation driven by brane motion and noncanonical kinetic dynamics~\cite{KKLMMT:2003,DBI:2004}. K\"ahler moduli, racetrack potentials, and fibre directions provide complementary constructions~\cite{ConlonQuevedo:2006,Racetrack:2004,Fibre:2009}. Across these constructions, the tensor prediction depends on the field direction selected after moduli stabilization. Pole inflation maps inverse powers of a boundary coordinate into canonical exponentials~\cite{Kallosh:2013,KalloshLindeRoest:2013,Roest:2015}. Limiting Hodge theory assigns the pole residue through a discrete degree $d$~\cite{Schmid:1973,CattaniKaplanSchmid:1986}. Integral three form flux contracts the period vector into the Gukov Vafa Witten(GVW) superpotential~\cite{Gukov:2000,GKP:2002}. Flux vacua and their stationary branches are constrained by the integral lattice and the tadpole bound~\cite{DenefDouglas:2004,Denef:2008}. Geometry determines the available field distances, whereas flux stationarity selects the distance traversed by the background. The tensor coefficient inherits $d$ when the logarithmic saxion becomes the adiabatic direction.

Axion monodromy makes the branch orientation explicit in both brane and flux realizations~\cite{Silverstein:2008sg,McAllister:2010,Grimm:2014}. F term constructions relate the monodromy potential to integral flux data and to the backreaction of stabilized fields~\cite{Marchesano:2014,Blumenhagen:2014,Hebecker:2014}. Saxion response can flatten the axion potential before the observables are evaluated~\cite{Flauger:2010,Dong:2011,Landete:2018}. Backreacted field ranges provide the associated geometric constraint~\cite{BaumePalti:2016}. Lanza and Westphal studied this response at finite large complex structure distance, where polynomial nilpotent orbit terms remain relevant and instanton corrections are exponentially suppressed~\cite{Lanza:2025}. In the large complex structure example used here, the stationary saxion grows linearly at large axion displacement. The valley potential changes from $a^3$ asymptotically toward an approximately quadratic penumbral regime, and the normalized complex structure gradient approaches $\sqrt6$ at the type IV boundary~\cite{Lanza:2025}. The uplift contribution can have a small normalized gradient near its de Sitter minimum, while the late potential remains above the strong gradient criterion. Since $\partial_sV=0$ along the stationary valley, the axion remains tangent to the trajectory. The saxion follows the axion displacement, and the axionic path determines the canonical inflaton slope.

The present construction eliminates the monodromy invariant coordinate $X=e-ma$ before projecting onto the adiabatic direction. The stationary profile $X_v(s)\propto s^{-\nu}$ places $X$ along the massive normal direction and leaves the saxion tangent to the valley. The canonical distance $\varphi=\sqrt{d/2}\log s$ then maps the leading inverse saxion term to an exponential plateau. For the type IV mirror quintic branch,
\begin{equation}
\gamma_{\rm br}\simeq {\frac{\sqrt{d/2}}{qN_*}},
\qquad
{\frac{\gamma_{\rm LCS}}{\gamma_{\rm br}}}\simeq2qN_*,
\label{eq:intro_gamma_ratio}
\end{equation}
with $d=3$ and $\gamma_{\rm LCS}=\sqrt6$. The Hodge degree cancels from the leading ratio. A common boundary metric therefore supports distinct canonical slopes because the stationary equations assign different coordinates to the tangent and normal directions. Once flux stationarity aligns the saxion with the adiabatic tangent, the metric degree $d$ enters the tensor prediction.

The integral mirror quintic period basis determines the Hodge norm and the polynomial flux contraction~\cite{Candelas:1991,Hosono:1995}. Boundedness of the active type IV no scale sector removes the flux components paired with the cubic and quadratic periods and leaves an integral representative with $N_{\rm flux}=0$. We then solve $\partial_XV=0$, substitute $X_v(s)$ into $V$, and evaluate the slow roll observables. No scale breaking terms generate the positive plateau energy and its leading finite distance deformation. A higher inverse power ends slow roll. The Becker Becker Haack Louis(BBHL) correction to the K\"ahler geometry is combined with the Kachru Kallosh Linde Trivedi (KKLT)nonperturbative sector to calculate the volume response, barrier, and Schur correction~\cite{BBHL:2002,KKLT:2003,LVS:2005}. Covariant evolution tests attraction toward the moving valley and quantifies the bending correction~\cite{Gordon:2000,GrootNibbelink:2000,Achucarro:2010}. At $N_*=55$, the official ACT DR6.02 tensor posterior determines the central reduced coefficients. The official running product contributes the correlated weight at the common pivot~\cite{ACTDR6LCDM,ACTDR6Extended,ACTDR6Chains}.

The active one parameter type IV branch is followed through its K\"ahler response and effective field theory hierarchy across the CMB interval. At the ACT inferred pivot $s_*=81.39$, the $s^{-2}$ deformation is $2.35\times10^{-2}$ of the leading $s^{-1}$ falloff. The exit operator contributes $1.53\times10^{-6}$ at the pivot and becomes comparable to the leading falloff near the end of slow roll. The instanton factor is approximately $10^{-222}$, so polynomial terms govern the finite distance correction throughout the observable interval. After saxion alignment, the Hodge degree determines the leading tensor coefficient, the Laurent coefficients determine the scalar curvature, and the endpoint coefficient determines $s_{\rm end}$. These quantities follow from $\partial_XV=0$, the normal mass hierarchy, the reduced potential $U(s)=V[s,X_v(s)]$, and posterior propagation of the resulting observables.

\inlinewidefigure{0.88}{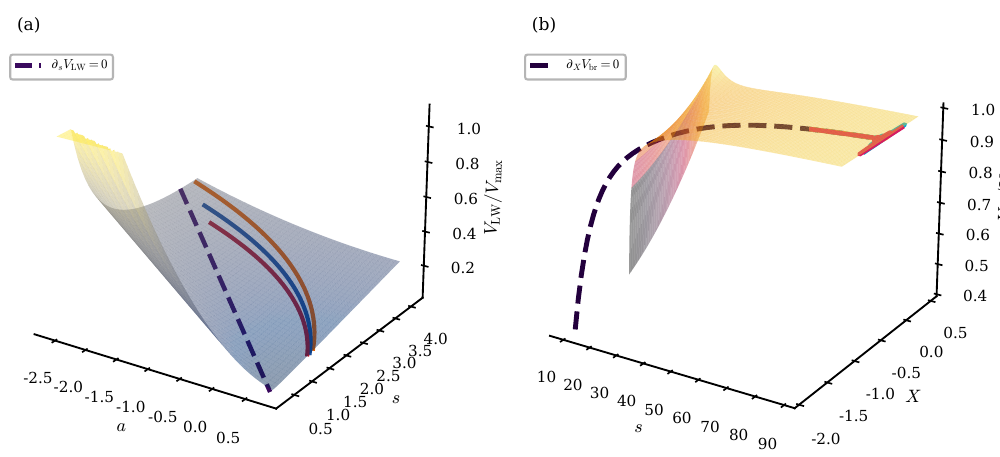}{Three dimensional comparison between the analytic axion led toy valley in Sec.~5.3 of Ref.~\cite{Lanza:2025} and the saxion led displaced branch. Panel (a) uses $p=q=m=e=1$, so $s_v(a)=|1-a|$ follows from $\partial_sV_{\rm toy}=0$. Panel (b) uses the $d=3$ branch with $X_v(s)=-20/s$ from $\partial_XV_{\rm br}=0$. Translucent surfaces display separately normalized potentials and compare the valley morphology. Dashed curves mark stationary valleys, and solid curves solve the covariant field equations. The toy surface visualizes the axionic orientation. The mirror quintic large complex structure coefficient used in the quantitative comparison is shown separately in Figure~\ref{fig:branch_diag}.}{fig:lw3dcomparison}

\inlinewidefigure{0.74}{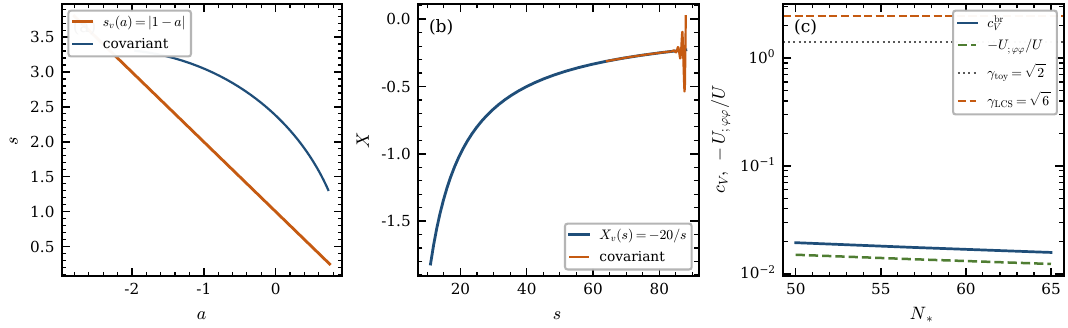}{Branch comparison. The left panel compares the stationary profile $s_v(a)=|1-a|$ of the analytic $p=q=m=e=1$ toy in Ref.~\cite{Lanza:2025} with its covariant solution. The middle panel compares $X_v(s)=-20/s$ with the displaced covariant solution. The right panel displays the gradient and curvature of the $d=3$ inflationary branch together with $\gamma_{\rm toy}=\sqrt2$ for the analytic $d=1$ toy and $\gamma_{\rm LCS}=\sqrt6$ for the type IV large complex structure example in Ref.~\cite{Lanza:2025}.}{fig:branch_diag}

\inlinewidefigure{0.74}{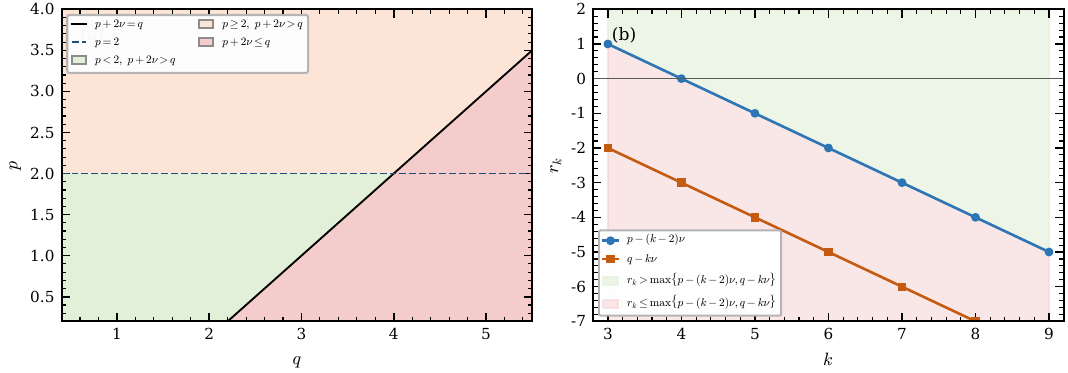}{Local control criterion and higher operator extension. The left panel displays the plateau and single clock domains in the $(q,p)$ plane for $\nu=1$, extending the local result of Ref.~\cite{PirzadaLi:2026}. The green domain satisfies both the plateau and asymptotic decoupling conditions. The right panel displays the lower saxion weights required for period derived $X^k$ operators to remain subleading on the original valley. An operator below the boundary requires a new dominant balance solution for $X_v(s)$.}{fig:theorem}

\section{Notation}\label{sec:defs}

We isolate one active complex structure coordinate
\begin{equation}
z=a+is.
\end{equation}
The coordinate \(a\) is axionic, and \(s>0\) is the saxion. The degree $d$ of the leading Hodge norm at the chosen boundary determines the boundary metric and the canonical saxionic variable. In the one parameter degeneration classification, $d=1,2,3$ distinguish type II, type III, and type IV growth sectors~\cite{CattaniKaplanSchmid:1986,GrimmPaltiValenzuela:2018}. The analytic reference class $d=1$ and the mirror quintic class $d=3$ consequently have different tensor normalizations. The monodromy charge is denoted by $m$, and $e$ labels the local branch position. The invariant heavy coordinate is
\begin{equation}
X=e-ma .
\end{equation}
Extremizing the transverse heavy direction, $\partial_XV=0$, yields the attractor valley $X_v(s)$. The powers $(q,p,\nu)$ specify, respectively, the leading saxion falloff in the reduced potential, the saxion weight of the quadratic heavy operator, and the saxion power controlling the branch displacement. Coefficients $c_q$, $A_p$ and $B_{p+\nu}$ denote the leading uplift, quadratic and odd flux terms in the local expansion. The number of e folds between pivot exit and the end of slow roll is $N_*$. The observables are the scalar tilt $n_s$, tensor ratio $r$ and running $\alpha_s$. The K\"ahler modulus is $T=\tau+i\rho$, the axiodilaton is $S=C_0+i/g_s$, and all potentials are expressed in reduced Planck units unless a scale is displayed. The period and flux superpotential conventions follow Refs.~\cite{Gukov:2000,GKP:2002}. The volume sector follows the nonperturbative and alpha prime corrected supergravity conventions of Refs.~\cite{KKLT:2003,BBHL:2002}.

\section{Periods and branches}

\subsection{Boundary metric}

Let $z=a+is$ be a complex structure coordinate with monodromy $z\mapsto z+1$. Near a unipotent boundary the period vector has a nilpotent orbit expansion with exponentially suppressed corrections~\cite{Schmid:1973,CattaniKaplanSchmid:1986}. Regular singular period equations provide the local analytic setting for this expansion~\cite{Deligne:1970}.
\begin{equation}
\Pi(z)=\exp(zN)\,\bm a_0+\sum_{n\ge1}\ee^{2\pi i n z}\bm a_n,
\label{eq:nilpotent}
\end{equation}
The nilpotent matrix \(N\) is the logarithm of the monodromy matrix. The Hodge norm
\begin{equation}
\mathcal H(z,\bar z)=i\bar\Pi^{T}\Sigma\Pi
\end{equation}
determines the complex structure K\"ahler potential
\begin{equation}
K_{\cs}=-\log \mathcal H .
\end{equation}
When the leading Hodge norm scales as
\begin{equation}
\mathcal H(s)=\kappa_d s^d\left[1+\Order(s^{-3},\ee^{-2\pi s})\right],
\label{eq:HodgeDegree}
\end{equation}
the K\"ahler potential necessarily reduces to
\begin{equation}
K_{\cs}=-d\log s+K_0+\Order(s^{-3},\ee^{-2\pi s}).
\end{equation}
Since $s=(z-\bar z)/(2i)$, one has
\begin{equation}
K_{z\bar z}=\partial_z\partial_{\bar z}K_{\cs}={\frac{d}{4s^2}}+\cdots .
\end{equation}
Enforcing the real field kinetic normalization $\frac12G_{IJ}\partial\phi^I\partial\phi^J=K_{z\bar z}\partial z\partial\bar z$ converts the Hodge norm into the hyperbolic boundary metric
\begin{equation}
\dd\ell^2={\frac{d}{2s^2}}(\dd a^2+\dd s^2)+\cdots .
\label{eq:dmetric}
\end{equation}
The integer \(d\) is the degree of the leading Hodge norm at the boundary and determines the canonical normalization. A one parameter Calabi Yau threefold at a maximally unipotent large complex structure boundary has \(d=3\). The analytic reference model with \(d=1\) belongs to a different boundary class. Equation~\eqref{eq:dmetric} connects topology to observables. After branch displacement makes the trajectory saxionic, the exponential slope is \(q\sqrt{2/d}\) and the leading tensor amplitude scales as \(r\simeq4d/(q^2N_*^2)\). The tensor amplitude retains the boundary class dependence through \(d\), while the leading scalar tilt follows from the logarithmic pole common to the boundary metric. The degree \(d\) is a quantized pole parameter determined by Hodge theory. Continuous supergravity coefficients leave the Hodge degree unchanged.

\subsection{Flux branch coordinate}

The type IIB flux superpotential is the Gukov Vafa Witten period contraction~\cite{Gukov:2000}
\begin{equation}
W_{\rm flux}=\int(F_3-SH_3)\wedge\Omega=(f-Sh)^T\Sigma\Pi(z),
\label{eq:gvw}
\end{equation}
where $f$ and $h$ are integral flux vectors, $S=C_0+i/g_s$ is the axiodilaton, and $\Sigma$ is the symplectic intersection form. The no scale complex structure potential follows from the type IIB flux effective theory~\cite{GKP:2002,Grana:2006}.
\begin{align}
V_{\cs}=&\,e^{K_{\cs}+K_S}\left(K^{z\bar z}|D_zW_{\rm flux}|^2+K^{S\bar S}|D_SW_{\rm flux}|^2\right),
\label{eq:Vcsfull}
\end{align}
The axiodilaton K\"ahler potential is \(K_S=-\log[-i(S-\bar S)]\). The axion enters through the polynomial part of $\Pi(z)$ and through instantons. The monodromy invariant branch coordinate is
\begin{equation}
X=e-ma,
\label{eq:Xdef}
\end{equation}
The integer \(m\) is the local monodromy charge, and \(e\) labels the branch. Using $X$ for the reduction retains the transverse branch explicitly.

Projecting the GVW superpotential onto the branch coordinate $X$ organizes the scalar potential into a hierarchical polynomial expansion~\cite{Marchesano:2014,Blumenhagen:2014,Hebecker:2014}
\begin{equation}
V(X,s)=C_0(s)+C_1(s)X+C_2(s)X^2+\sum_{k\ge3}C_k(s)X^k.
\label{eq:Cexpansion}
\end{equation}
We impose the penumbral ordering, over the interval of interest,
\begin{align}
C_0(s)&=V_0-c_qV_0s^{-q}+\tilde c_2V_0s^{-2q}+\cdots,\label{eq:C0hier}\\
C_1(s)&=B_{p+\nu}s^{-(p+\nu)}+\cdots,\label{eq:C1hier}\\
C_2(s)&=A_ps^{-p}+\cdots,\label{eq:C2hier}
\end{align}
with $V_0>0$, $c_q>0$ and $A_p>0$. The power $q$ is the leading uplift falloff, $p$ is the heavy Hessian power and $\nu$ is the displacement exponent. The ordering is local. In a completed orientifold, every coefficient is determined by the interplay between the period vector, the integral flux lattice, the axiodilaton value and the stabilized spectator sector.

\subsection{Axion led and saxion led branches}

Branch displacement determines the adiabatic direction before any single field projection. Ref.~\cite{Lanza:2025} and the present construction both use a unipotent coordinate \(z=a+is\), a hyperbolic boundary metric, and polynomial flux data. The reductions differ in the coordinate eliminated by its stationary equation.

The analytic family in Sec.~5.3 of Ref.~\cite{Lanza:2025} is
\begin{equation}
V_{\rm toy}(a,s)={\frac{(e-ma)^2}{2s^p}}+{\frac{m^2}{2}}s^q+V_0 .
\label{eq:VLWtoy}
\end{equation}
The saxion profile along the axion valley is
\begin{equation}
s_v(a)=\left({\frac{p(e-ma)^2}{q m^2}}\right)^{1/(p+q)} .
\label{eq:svLW}
\end{equation}
For \(p=q=m=e=1\), the profile is \(s_v=|1-a|\). The metric of the analytic toy has \(d=1\), and the stationary potential grows linearly at large \(|a|\). Since \(\partial_sV_{\rm toy}=0\) on the valley, the normal derivative drops out of the covariant norm. The remaining axionic derivative gives
\begin{equation}
\gamma_{\rm toy}^2=G^{aa}{\frac{(\partial_aV_{\rm toy})^2}{V_{\rm toy}^2}},
\qquad
G^{aa}={\frac{2s^2}{d}},
\label{eq:gammaLW}
\end{equation}
and therefore
\begin{equation}
\gamma_{\rm toy}\longrightarrow\sqrt2 .
\label{eq:gammaLWnum}
\end{equation}
The toy family isolates the geometric mechanism behind polynomial flattening without identifying the coefficient of the separate mirror quintic example.

The large complex structure model in Sec.~5.1 of Ref.~\cite{Lanza:2025} provides the type IV comparison. Its stationary saxion grows linearly with the axion at large displacement, while the valley potential scales as $a^3$ before softening toward $a^2$ in the penumbral interval~\cite{Lanza:2025}. The associated uplift gradient approaches
\begin{equation}
\gamma_{\rm LCS}=\sqrt6
\label{eq:gammaLCS}
\end{equation}
near the boundary~\cite{Lanza:2025}. The uplift gradient becomes arbitrarily small near the de Sitter minimum, whereas the late gradient remains above the strong de Sitter bound along the axion valley~\cite{Lanza:2025}. Polynomial backreaction therefore produces genuine flattening without making the asymptotic axion tangent a slow roll direction. Equation~\eqref{eq:gammaLCS} provides the boundary coefficient for a trajectory level comparison. On the reference axion valley the saxion derivative vanishes, leaving the normalized axion slope. On the displaced branch the heavy invariant derivative vanishes, leaving the normalized saxion slope. The comparison therefore holds the type IV metric constant and changes the stationary tangent.

The displaced construction imposes the stationary condition on
\begin{equation}
X=e-ma
\end{equation}
before the adiabatic projection. The leading branch operators yield
\begin{equation}
X_v(s)=-{\frac{B_{p+\nu}}{2A_p}}s^{-\nu},
\label{eq:Xvcomparison}
\end{equation}
so \(X\) follows the massive normal direction while \(s\) parametrizes the trajectory. The induced canonical distance is
\begin{equation}
\varphi_{\rm br}\simeq \sqrt{{\frac{d}{2}}}\log s.
\end{equation}
The coordinate exchange changes the canonical derivative without changing the type IV metric. At leading order in the e fold expansion,
\begin{equation}
\gamma_{\rm br}=\sqrt{2\epsilon_*}
\simeq {\frac{\sqrt{d/2}}{qN_*}}.
\label{eq:gammabrleading}
\end{equation}
For the type IV value $d=3$, combining Eqs.~\eqref{eq:gammaLCS} and \eqref{eq:gammabrleading} gives
\begin{equation}
{\frac{\gamma_{\rm LCS}}{\gamma_{\rm br}}}
\simeq2qN_*.
\label{eq:gammaratio}
\end{equation}
The Hodge degree cancels in the type IV comparison. The slope hierarchy measures the duration of the saxionic attractor while the boundary class remains unchanged. For \(q=1\) and \(N_*=55\), Eq.~\eqref{eq:gammaratio} gives \(110\). Direct integration of the ACT inferred benchmark gives
\begin{equation}
\gamma_{\rm br}=\sqrt{{\frac{r_*}{8}}}=1.80\times10^{-2},
\qquad
{\frac{\sqrt6}{\gamma_{\rm br}}}=135.9,
\label{eq:gammabr}
\end{equation}
The difference from the leading value follows from the finite distance $s^{-2}$ curvature term and the full relation between $s_*$, $s_{\rm end}$, and $N_*$.

\begin{center}
\scriptsize
\begin{tabular}{@{}lll@{}}
\hline\hline
 & LCS valley of Ref.~\cite{Lanza:2025} & present branch\\
\hline
light coordinate & \(a\) & \(s\)\\
stationary response & \(s_v(a)\) & \(X_v(s)\)\\
stationary equation & \(\partial_sV=0\) & \(\partial_XV=0\)\\
canonical distance & \(\log s_v(a)\) & \(\log s\)\\
asymptotic slope & \(\sqrt6\) & \(\sqrt{d/2}/(qN_*)\)\\
physical role & polynomial flattening & logarithmic slow roll\\
\hline\hline
\end{tabular}
\end{center}

Figure~\ref{fig:lw3dcomparison} uses the analytic toy of Ref.~\cite{Lanza:2025} because its closed stationary profile permits a direct covariant three dimensional comparison. The reference parameters \(p=q=m=e=1\) give \(s_v(a)=|1-a|\) with no auxiliary scale. The displaced panel uses
\begin{equation}
\widehat V_{\rm br}(s,X)=U(s)+{\frac{35d}{12}}s^{-2}\left(X+{\frac{20}{s}}\right)^2,
\qquad d=3.
\label{eq:Vbr3D}
\end{equation}
The coefficient \(20=B_3/(2A_2)\) determines the stationary displacement for \(p=2\) and \(\nu=1\). The factor \(35d/12\) gives \(m_N^2/H^2=35\) after using \(G^{XX}=2s^2/d\) and \(H^2\simeq V_0/3\). One coefficient locates the valley, and the other measures its transverse stiffness. The comparison therefore separates branch orientation from normal stabilization before the compact coefficient map is introduced.

\section{Branch plateau theorem}

\subsection{Stationary branch and canonical map}

The leading polynomial operators isolate the branch mechanism.
\begin{equation}
V(X,s)=V_0-c_qV_0s^{-q}+A_ps^{-p}X^2+B_{p+\nu}s^{-(p+\nu)}X+\cdots .
\label{eq:localV}
\end{equation}
The leading correction is the inverse saxion uplift inherited from the boundary expansion, the quadratic term is the heavy invariant mass, and the odd term is the branch displacing flux operator. The ordering between these powers determines both the plateau and the heavy field gap. Extremizing in the heavy coordinate yields
\begin{equation}
0=\partial_XV=2A_ps^{-p}X+B_{p+\nu}s^{-(p+\nu)}+\cdots,
\end{equation}
so the displaced branch is
\begin{equation}
X_v(s)=-{\frac{B_{p+\nu}}{2A_p}}s^{-\nu}+\Order(s^{-\nu-1}).
\label{eq:Xv}
\end{equation}
Since $a=(e-X)/m$,
\begin{equation}
{\frac{\dd a_v}{\dd s}}=-{\frac{1}{m}}{\frac{\dd X_v}{\dd s}}=\Order(s^{-\nu-1}).
\label{eq:avslope}
\end{equation}
The induced line element on the branch is
\begin{align}
\dd\varphi^2&={\frac{d}{2s^2}}\left(1+a_v'(s)^2\right)\dd s^2
={\frac{d}{2s^2}}\left[1+\Order(s^{-2\nu-2})\right]\dd s^2.
\end{align}
Integration of the induced line element yields
\begin{equation}
\varphi=\sqrt{\frac{d}{2}}\log s+\Order(s^{-2\nu-2}),
\label{eq:canon}
\end{equation}
so inverse powers of $s$ become plateau exponentials,
\begin{equation}
s^{-n}=\exp\left(-n\sqrt{\frac{2}{d}}\,\varphi\right)\left[1+\Order\left(\ee^{-(2\nu+2)\sqrt{2/d}\varphi}\right)\right].
\end{equation}
The exponential slope is the product of two independent data. The saxion weight \(n\) originates in the flux expansion, while \(\sqrt{2/d}\) originates in the geodesic conversion imposed by the Hodge norm. A larger Hodge degree stretches canonical distance and lowers the exponential slope while leaving the Laurent power in \(s\) unchanged. The separation explains why \(d\) survives in the tensor amplitude while local flux coefficients enter subleading observables.

Substituting Eq.~\eqref{eq:Xv} into Eq.~\eqref{eq:localV} yields
\begin{equation}
U(s)=V_0-c_qV_0s^{-q}-{\frac{B_{p+\nu}^2}{4A_p}}s^{-(p+2\nu)}+\cdots .
\label{eq:Us}
\end{equation}
The uplift plateau dominates the branch energy when
\begin{equation}
\Delta=p+2\nu-q>0.
\label{eq:Delta}
\end{equation}
The inequality compares the energy gained from transverse relaxation with the leading approach to the plateau. The ratio of the two contributions scales as \(s^{-\Delta}\). Positive \(\Delta\) allows branch displacement to orient the trajectory while the displacement energy decays relative to the plateau falloff. Integrating out \(X\) changes the tangent direction before changing the leading energy. The canonical potential is then
\begin{equation}
U(\varphi)=V_0\left[1-c_q\ee^{-q\sqrt{2/d}\varphi}+\cdots\right].
\end{equation}
The odd flux term changes the field measuring distance along the valley. The reduced trajectory inherits the logarithmic saxion distance, while the branch coordinate remains the heavy invariant field. Equation~\eqref{eq:canon} is the geometric engine of the construction because the logarithmic field redefinition acts as a field space conformal map. By identifying the canonical field with the logarithmic saxion distance, the hyperbolic boundary metric separates inflationary flatness from the detailed polynomial structure of the flux superpotential. Any inverse saxion falloff becomes an exponential plateau whose decay rate follows from the Hodge norm. In the penumbral control regime the instanton terms remain exponentially suppressed, while the lightest tower remains above the Hubble scale over the CMB interval. Under these assumptions, the Hodge degree $d$ and the saxion weight $q$ determine the flatness. The heavy response appears through the subleading power \(p+2\nu\) and through the entropy mass.

Figure~\ref{fig:branch_diag} separates the analytic toy normalization from the mirror quintic large complex structure result. The toy profile \(s_v(a)=|1-a|\) approaches \(\gamma_{\rm toy}=\sqrt2\). The type IV large complex structure model approaches \(\gamma_{\rm LCS}=\sqrt6\), while the displaced \(d=3\) background reaches \(\gamma_{\rm br}=1.81\times10^{-2}\). The three curves distinguish polynomial flattening, boundary normalization, and logarithmic slow roll.

\subsection{\texorpdfstring{Large-$N_*$ predictions}{Large N predictions}}

Let $a_q=q\sqrt{2/d}$. At leading order,
\begin{equation}
{\frac{U_{,\varphi}}{U}}=a_qc_q\ee^{-a_q\varphi}+\cdots,
\qquad
{\frac{U_{,\varphi\varphi}}{U}}=-a_q^2c_q\ee^{-a_q\varphi}+\cdots .
\end{equation}
The e fold integral is
\begin{equation}
N_*\simeq\int^{\varphi_*}{\frac{U}{U_{,\varphi}}}\dd\varphi={\frac{\ee^{a_q\varphi_*}}{a_q^2c_q}}+\cdots .
\end{equation}
Eliminating $\varphi_*$ yields
\begin{equation}
\epsilon_*={\frac{1}{2a_q^2N_*^2}}+\Order(N_*^{-3})={\frac{d}{4q^2N_*^2}}+\Order(N_*^{-3}),
\label{eq:epslarge}
\end{equation}
\begin{equation}
\eta_*=-{\frac{1}{N_*}}+\Order(N_*^{-2}).
\end{equation}
The large $N_*$ observables are
\begin{align}
n_s&=1-{\frac{2}{N_*}}+\Order(N_*^{-2}),\label{eq:nslarge}\\
r&={\frac{4d}{q^2N_*^2}}+\Order(N_*^{-3}),\label{eq:rlarge}\\
\alpha_s&=-{\frac{2}{N_*^2}}+\Order(N_*^{-3}).\label{eq:alphalarge}
\end{align}
The coefficient \(c_q\) cancels after the horizon position is eliminated in favor of \(N_*\). The cancellation explains the leading universality of \(n_s\) and \(\alpha_s\). The tensor amplitude retains \(d\) because canonical distance depends on the residue of the boundary pole. Comparison with \(r=12\alpha/N_*^2\) yields
\begin{equation}
\alpha_{\rm pole}={\frac{d}{3q^2}}.
\label{eq:alphapole}
\end{equation}
Phenomenological attractor models treat \(\alpha\) as a continuous supergravity parameter~\cite{Kallosh:2013,KalloshLindeRoest:2013,Roest:2015}. A Calabi Yau boundary replaces the continuous freedom with the discrete Hodge degree \(d\), while flux data reappear through corrections beyond the leading pole limit.

\subsection{Entropy mass}

In $(X,s)$ variables the boundary metric is
\begin{equation}
\dd\ell^2={\frac{d}{2s^2}}\left({\frac{\dd X^2}{m^2}}+\dd s^2\right).
\label{eq:metricXs}
\end{equation}
Define $y=X_v'(s)/m$. Unit tangent and normal vectors are
\begin{align}
T^I&={\frac{\sqrt{2/d}\,s}{\sqrt{1+y^2}}}(my,1),\label{eq:Tvec}\\
N^I&={\frac{\sqrt{2/d}\,s}{\sqrt{1+y^2}}}(m,-y).\label{eq:Nvec}
\end{align}
The tangent and normal vectors obey $G_{IJ}T^IT^J=G_{IJ}N^IN^J=1$ and $G_{IJ}T^IN^J=0$. The covariant Hessian is
\begin{equation}
V_{;IJ}=\partial_I\partial_JV-\Gamma^K{}_{IJ}\partial_KV.
\label{eq:covHess}
\end{equation}
Because $y=\Order(s^{-\nu-1})$, the normal aligns with $X$ and
\begin{equation}
N^IN^JV_{;IJ}={\frac{2m^2s^2}{d}}\left(2A_ps^{-p}\right)\left[1+\Order(s^{-\nu-1})\right].
\end{equation}
The normal Hessian, perturbative entropy mass, and heavy denominator entering the adiabatic EFT are distinct quantities. We define
\begin{align}
m_N^2&=V_{;NN},\label{eq:mNdef}\\
m_{\rm iso}^2&=V_{;NN}+3\Omega^2+\epsilon H^2R_{\rm fs},\label{eq:msdef}\\
M_{\rm eff}^2&=m_N^2-\Omega^2+\epsilon H^2R_{\rm fs},\label{eq:Meffdef}
\end{align}
The bending rate is \(\Omega=|D_tT^I|\), and Eq.~\eqref{eq:dmetric} has field space curvature \(R_{\rm fs}=-4/d\). The leading normal Hessian yields
\begin{equation}
{\frac{m_{\rm iso}^2}{H^2}}={\frac{12A_pm^2}{dV_0}}s^{2-p}+\Order(\epsilon,\Omega^2/H^2).
\label{eq:mxscale}
\end{equation}
For \(p<2\), the entropy mode is parametrically heavy toward the boundary. The mass ratio decreases as the saxion rolls inward, so the relevant control quantity is \(\min_{\rm CMB}(m_{\rm iso}^2/H^2)\). The marginal power \(p=2\) removes the saxion scaling and places the entire decoupling test in the coefficient \(12A_pm^2/(dV_0)\). Branch ordering links two physical effects. The odd flux operator rotates the tangent direction, while the quadratic coefficient determines how rapidly the normal displacement decays.

\section{Local control theorem}

The local result of Ref.~\cite{PirzadaLi:2026} classifies the leading displaced valley by $(q,p,\nu)$. The plateau condition is $p+2\nu>q$, and the entropy mode is parametrically heavy for $p<2$, with $p=2$ controlled by the coefficient. The present construction connects the weights to period derived coefficients and tests preservation of the leading balance under higher branch operators.

Let a higher operator be
\begin{equation}
\delta V_k=D_ks^{-r_k}X^k,
\qquad k\ge3.
\end{equation}
On the leading valley $X_v\sim s^{-\nu}$, the operator contribution to the branch equation scales as
\begin{equation}
\partial_X\delta V_k\sim s^{-[r_k+(k-1)\nu]}.
\end{equation}
The leading branch equation and normal Hessian remain unchanged when
\begin{equation}
r_k+(k-2)\nu>p,
\label{eq:harmless1}
\end{equation}
and the operator contribution to the reduced energy remains beyond the plateau term when
\begin{equation}
r_k+k\nu>q.
\label{eq:harmless2}
\end{equation}
Inequalities~\eqref{eq:harmless1} and \eqref{eq:harmless2} are sufficient irrelevance conditions. The two inequalities preserve the original stationary branch, entropy scaling, and plateau exponent under the added operator.

Failure of either condition invalidates the original displacement exponent. Terms becoming comparable in $\partial_XV$ define a new dominant balance and a new branch exponent. Besides the original quadratic and source balance $\nu_{QS}=\nu$, the candidate exponents are
\begin{align}
\nu_{QH}&={\frac{p-r_k}{k-2}},\label{eq:nuQH}\\
\nu_{SH}&={\frac{p+\nu-r_k}{k-1}},\label{eq:nuSH}
\end{align}
for quadratic and higher operator and source and higher operator balance, respectively. A candidate defines a branch when the associated coefficient equation has a real stable solution and omitted terms remain asymptotically subleading on the candidate solution. The reduced energy and entropy mass are then recalculated with the resulting $\nu_{\rm eff}$. Figure~\ref{fig:theorem} displays the sufficient irrelevance region and the corresponding lower saxion weights. The formulation preserves the derived leading branch and identifies the separate dominant balance problem arising when a higher operator becomes active.

Hodge theory determines the boundary metric, the stationary coefficient map determines the branch expansion, and the K\"ahler sector determines the volume response. A higher operator enters $\partial_XV$ before the reduced potential because it can change the stationary valley while its direct contribution to the reduced energy remains small. The corrected stationary solution determines the normal Hessian, and substitution into $V$ then gives the reduced potential.

\section{Mirror quintic coefficient map}

\subsection{Picard Fuchs monodromy}

The Greene Plesser mirror construction and the standard mirror quintic family define the quintic hypersurface~\cite{GreenePlesser:1990,Candelas:1991}
\begin{equation}
P_\psi=x_1^5+x_2^5+x_3^5+x_4^5+x_5^5-5\psi x_1x_2x_3x_4x_5=0
\label{eq:quinticpoly}
\end{equation}
in $\mathbb P^4$ after quotienting by the Greene Plesser phase group \(G\simeq(\mathbb Z_5)^3\) preserving \(P_\psi\) and the holomorphic three form. The invariant complex structure coordinate is
\begin{equation}
z=(5\psi)^{-5},
\end{equation}
so the large complex structure point is $z=0$ and the flat coordinate $t$ is the logarithmic period ratio. The orbifold presentation determines the monodromy action before any inflationary ansatz is introduced. The relation $t\mapsto t+1$ is the maximally unipotent monodromy generated by the logarithmic solutions of the Picard Fuchs equation.

The one parameter mirror quintic arises from the Greene Plesser quotient and its toric mirror construction~\cite{GreenePlesser:1990,Batyrev:1994}. Its periods obey the Picard Fuchs operator~\cite{Candelas:1991,CoxKatz}
\begin{equation}
\mathcal L=\theta^4-5z(5\theta+1)(5\theta+2)(5\theta+3)(5\theta+4),
\qquad \theta=z{\frac{\dd}{\dd z}}.
\label{eq:PF}
\end{equation}
The large complex structure flat coordinate is
\begin{equation}
t={\frac{1}{2\pi i}}\log z+\Order(z),
\qquad t=a+is,
\end{equation}
with $t\mapsto t+1$. In the symplectic large complex structure basis used in mirror quintic period calculations~\cite{Candelas:1991,Morrison:1993,CoxKatz},
\begin{equation}
\Pi(t)=
\begin{pmatrix}
1\\[2pt]
t\\[2pt]
{\frac{5}{6}}t^3+{\frac{25}{12}}t+\xi_0\\[2pt]
-{\frac{5}{2}}t^2-{\frac{11}{2}}t+{\frac{25}{12}}
\end{pmatrix}
+\Order(\ee^{2\pi it}),
\label{eq:Pi}
\end{equation}
with the ordering $\Pi=(X^0,X^1,F_0,F_1)^T$. The Euler characteristic $\chi(X)=-200$ of the quintic mirror partner determines the constant term,
\begin{equation}
\xi_0=-{\frac{25i\,\zeta(3)}{\pi^3}}.
\label{eq:xi0}
\end{equation}
The instanton series is exponentially small over the benchmark interval. At $s_*=81.39$, $\ee^{-2\pi s_*}\simeq10^{-222}$, while the retained $s^{-2}$ correction is $2.35\times10^{-2}$ relative to the leading branch term. The CMB calculation probes the finite distance polynomial penumbra within the nilpotent orbit regime. Contracting the period vector with the symplectic form yields the leading Hodge norm
\begin{equation}
i\bar\Pi^{T}\Sigma\Pi={\frac{20}{3}}s^3+{\frac{2\zeta(3)|\chi|}{(2\pi)^3}}.
\label{eq:HodgeNormMQ}
\end{equation}
The nilpotent orbit relation contains no intermediate powers of \(s\) and no axionic dependence. The absence of \(a\) expresses the shift symmetry of the nilpotent orbit. The monodromy invariant combination \(X=e-ma\) is therefore the transverse branch coordinate. The constant \(2\zeta(3)|\chi(X)|/(2\pi)^3=1.9384\) is the complex structure \(\alpha'^3\) correction, and it is mirror to the BBHL term \(\hat\xi\) used in the K\"ahler sector in Eq.~\eqref{eq:BBHLK}. The complex structure prepotential uses $\chi(X)=-200$ of the quintic mirror partner, whereas BBHL uses $\chi(Y)=+200$ for the mirror quintic compactification. Mirror symmetry preserves the magnitude and reverses the sign, so the two corrections cannot be assigned one common Euler characteristic. The period vector also transforms under \(t\mapsto t+1\) by an integral symplectic matrix obeying \(T^{T}\Sigma T=\Sigma\), so the branch label \(e\) is a genuine monodromy orbit index and not a free parameter.
The cubic power is the geometric datum relevant for inflation and yields
\begin{equation}
K_{\cs}=-3\log s+K_0+\Order(s^{-3},\ee^{-2\pi s}),
\label{eq:Kcs}
\end{equation}
and identifies \(d=3\). The constant and inverse saxion terms deform the finite distance potential while leaving the kinetic pole unchanged. Consequently, the leading tensor coefficient is insensitive to the period normalization while the scalar curvature remains sensitive to subleading period and flux data. The monodromy is represented by an integral symplectic matrix $T=\ee^N$ acting on $\Pi$, with flat coordinate action $t\mapsto t+1$. The branch coordinate $X=e-ma$ is invariant under the combined monodromy $a\mapsto a+1$, $e\mapsto e+m$.

\subsection{Flux polynomial}

\begin{figure*}[!t]
\includegraphics[width=0.96\textwidth]{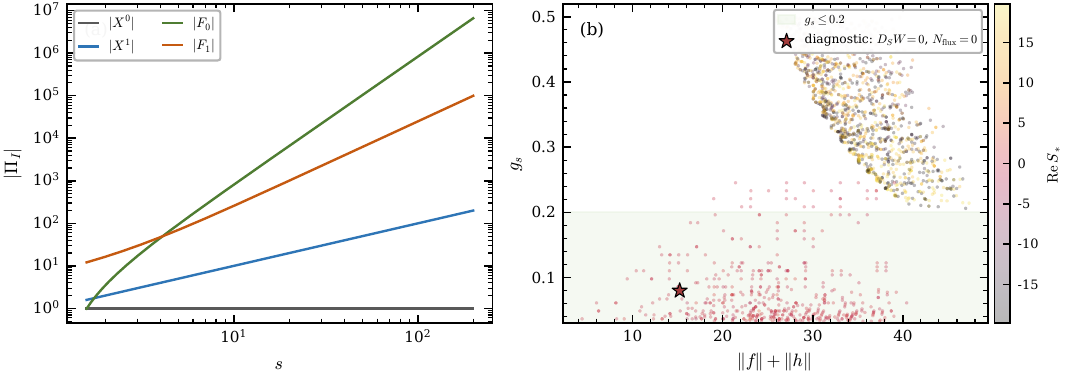}
\caption{Mirror quintic period growth and bounded active flux population. The left panel displays the nilpotent orbit hierarchy responsible for the type IV Hodge degree \(d=3\). The right panel samples integral vectors satisfying \(f_0=f_1=h_0=h_1=0\). Every sampled pair has \(N_{\rm flux}=0\). The green region marks \(g_s\leq0.2\), the color scale gives \(\operatorname{Re}S_*\), and the star marks the active representative in Eq.~\eqref{eq:fluxchoice}, which obeys $D_SW=0$ and $D_tW=-4$.}
\label{fig:periods}
\end{figure*}

With $\Pi=(X^0,X^1,F_0,F_1)^T$ and
\begin{equation}
\Sigma=\begin{pmatrix}0&I_2\\-I_2&0\end{pmatrix},
\end{equation}
the GVW superpotential of flux compactifications~\cite{Gukov:2000,GKP:2002} becomes
\begin{equation}
W_{\rm flux}(t,S)=w_0+w_1t+w_2t^2+w_3t^3+\Order(\ee^{2\pi it}).
\label{eq:Wpoly}
\end{equation}
For the bounded type IV active representative
\begin{equation}
f=(0,0,-1,-2),
\qquad
h=(0,0,13,0),
\label{eq:fluxchoice}
\end{equation}
the axiodilaton equation \(D_SW=0\) at the benchmark saxion gives
\begin{equation}
S_*=-0.076923+12.521682i,
\qquad
g_s=0.079861.
\label{eq:Sflux}
\end{equation}
The numerical residual satisfies \(|D_SW|<10^{-12}\). At the pivot, the active representative gives
\begin{equation}
W_*=325.563745i,
\qquad
D_tW_*=-4,
\label{eq:activeDtW}
\end{equation}
The nonzero complex structure derivative determines the active polynomial contribution to the stationary potential. In the adopted integral basis, the flux contractions have no cubic or quadratic terms,
\begin{align}
w_0&=162.781872i,
& w_1&=2,
& w_2&=w_3=0.
\label{eq:wcoeffs}
\end{align}
The symplectic pairing is
\begin{equation}
N_{\rm flux}=f^T\Sigma h=0.
\label{eq:Nflux25}
\end{equation}
The representative realizes the bounded active flux locus and leaves the D3 charge budget available for spectator stabilization. Its undressed active no scale energy at the pivot is $V_{\rm active}=1.56993\times10^{-3}$. Dressing by the corrected overall volume block at $\tau_0=100$ gives $1.60419\times10^{-9}=19.86V_*$. The active energy before and after volume dressing normalizes the bounded polynomial sector. The no scale breaking stationary terms determine the Laurent coefficients, displacement, and normal mass used below, and the ACT posterior selects the central point in the reduced coefficient space. Figure~\ref{fig:periods} displays the period growth and the weak coupling population of the bounded integral locus.

\subsection{Flux coefficient map}

Expanding the GVW F terms in the branch coordinate yields the microscopic coefficient map. The expansion determines how the period vector and integral flux contractions contribute to each local saxion weight~\cite{Gukov:2000,GKP:2002,DouglasKachru:2007}. Write
\begin{equation}
t=t_b-{\frac{X}{m}},
\qquad t_b={\frac{e}{m}}+is.
\end{equation}
At the boundary,
\begin{align}
K_t&={\frac{id}{2s}}, & K^{t\bar t}&={\frac{4s^2}{d}},\nonumber\\
K_S&=-{\frac{1}{S-\bar S}}, & K^{S\bar S}&=-(S-\bar S)^2.
\end{align}
For the polynomial $W(t)=\sum_{n=0}^{3}w_nt^n$ define
\begin{align}
\mathcal A_j(s)&={\frac{(-1/m)^j}{j!}}
\left[W^{(j+1)}(t_b)+{\frac{id}{2s}}W^{(j)}(t_b)\right],\label{eq:Aj}\\
\mathcal B_j(s)&={\frac{(-1/m)^j}{j!}}
\left[-H^{(j)}(t_b)-{\frac{1}{S-\bar S}}W^{(j)}(t_b)\right].\label{eq:Bj}
\end{align}
The polynomial \(H(t)=h^T\Sigma\Pi(t)\) contains the NS flux contraction, and \(W^{(j)}=\partial_t^jW\). The coefficients \(\mathcal A_j\) and \(\mathcal B_j\) are covariant Taylor coefficients of the complex structure and axiodilaton F terms along the invariant displacement. Then
\begin{align}
D_tW&=\sum_{j\ge0}\mathcal A_j(s)X^j,\label{eq:Dtseries}\\
D_SW&=\sum_{j\ge0}\mathcal B_j(s)X^j.\label{eq:DSseries}
\end{align}
Substituting into the no scale complex structure potential in Eq.~\eqref{eq:Vcsfull} yields the coefficient of each power of $X$
\begin{align}
C_\ell(s)&=e^{K_{\cs}+K_S}\Bigg[ {\frac{4s^2}{d}}\sum_{j+k=\ell}\mathcal A_j\overline{\mathcal A_k}
+K^{S\bar S}\sum_{j+k=\ell}\mathcal B_j\overline{\mathcal B_k}\Bigg].
\label{eq:Cell}
\end{align}
For a specified orientifold and stabilized heavy sector, Eqs.~\eqref{eq:Aj} through \eqref{eq:Cell} map the microscopic flux data into the branch coefficients governing the CMB dynamics. The Hermitian convolutions in Eq.~\eqref{eq:Cell} retain phase information from the integral flux vector, while the factors \(s^2/d\) and \(K^{S\bar S}\) encode the geometric weights of the two F term sectors. The Laurent powers originate in the boundary geometry, and the amplitudes retain arithmetic flux information. The leading branch data used in the theorem are extracted by
\begin{align}
C_0(s)&=V_0\left(1-{\frac{c_1}{s}}+{\frac{c_2}{s^2}}+\cdots\right),\label{eq:C0extract}\\
C_1(s)&=B_{p+\nu}s^{-(p+\nu)}+\cdots,\label{eq:C1extract}\\
C_2(s)&=A_ps^{-p}+\cdots.\label{eq:C2extract}
\end{align}
The compact arithmetic system is a Diophantine system in the integer fluxes.
\begin{equation}
(f,h)\in\Lambda_{\rm odd},\qquad f^T\Sigma h+N_{D3}=L,
\label{eq:tadpole}
\end{equation}
combined with Eqs.~\eqref{eq:Cell} through \eqref{eq:C2extract}. The no scale part of Eq.~\eqref{eq:Cell} is a positive Hermitian Gram form. The positivity converts asymptotic boundedness into linear restrictions on the integral fluxes. In the leading type IV Hodge approximation,
\begin{equation}
V_{\cs}^{(\mathrm{IV})}
={\frac{g_s}{10s}}|D_tW|^2
+{\frac{3}{10g_ss^3}}|D_SW|^2
+\Order(s^{-4}).
\label{eq:Vcsexact}
\end{equation}
Write \(F(t)=\sum F_nt^n\), \(H(t)=\sum H_nt^n\), and \(C_0^{(S)}=\operatorname{Re}S\). The coefficient of the leading \(s^3\) weight is
\begin{equation}
\alpha_3={\frac{3}{10}}
\left[g_s\bigl(F_3-C_0^{(S)}H_3\bigr)^2
+{\frac{H_3^2}{g_s}}\right].
\label{eq:alpha3}
\end{equation}
After \(F_3=H_3=0\), the coefficient of the remaining growing \(s\) weight is
\begin{equation}
\alpha_1={\frac{1}{10}}
\left[g_s\bigl(F_2-C_0^{(S)}H_2\bigr)^2
+{\frac{H_2^2}{g_s}}\right].
\label{eq:alpha1}
\end{equation}
Both coefficients are positive definite. A bounded active no scale sector therefore requires
\begin{equation}
F_3=H_3=F_2=H_2=0.
\label{eq:weightthm}
\end{equation}
For the symplectic ordering \(f=(f_0,f_1,f_2,f_3)\) used in Eq.~\eqref{eq:fluxchoice},
\begin{equation}
F_3={\frac56}f_0,
\qquad
F_2=-{\frac52}f_1,
\end{equation}
with corresponding relations for \(h\). Equation~\eqref{eq:weightthm} is equivalent to \(f_0=f_1=h_0=h_1=0\). The symplectic product then obeys
\begin{equation}
N_{\rm flux}=f_0h_2+f_1h_3-f_2h_0-f_3h_1=0.
\label{eq:Nfluxzero}
\end{equation}
The vanishing charge is a consequence of boundedness in the one parameter type IV active sector. Spectator fluxes on additional orientifold odd cycles enter separately in the total tadpole equation.

The weight decomposition identifies the origin of the plateau. After the growing active weights vanish, the no scale breaking K\"ahler, uplift, and cross coupling terms in Eq.~\eqref{eq:SUGRAV} generate the positive energy $V_0$ and the inverse saxion expansion. Flux arithmetic constrains the polynomial weights and branch orientation. Stationary reduction defines the Laurent coefficient map, and the ACT posterior selects the benchmark values of $c_1$, $c_2$, and $c_8$.

The table collects the geometric data, the bounded active flux representative, and the ACT inferred reduced coefficients used in the stabilization and CMB calculations.

\begin{center}
\scriptsize
\begin{ruledtabular}
\begin{tabular}{lc|lc}
\multicolumn{2}{c|}{period and active flux data} & \multicolumn{2}{c}{reduced branch data}\\
quantity & value & quantity & value\\
\hline
\(f\) & \((0,0,-1,-2)\) & \(c_1\) & \(1.8432\)\\
\(h\) & \((0,0,13,0)\) & \(c_2\) & \(3.5257\)\\
\(S_*\) & \(-0.076923+12.521682i\) & \(c_8\) & \(6.6603\times10^7\)\\
\(N_{\rm flux}\) & \(0\) & \(p\) & \(2\)\\
\(w_0\) & \(162.781872i\) & \(q\) & \(1\)\\
\(w_1\) & \(2\) & \(\nu\) & \(1\)\\
\(w_2,w_3\) & \(0,0\) & \(m_X^2/H^2\) & \(35\)\\
\(W_*,D_tW_*\) & \(325.563745i,-4\) & \(c_4^{\rm net}\) & \(0\)\\
\(d\) & \(3\) & \(\hat\xi/(2\mathcal V)\) & \(-1.074\times10^{-2}\)\\
\end{tabular}
\end{ruledtabular}
\end{center}

\subsection{Stabilized branch extraction}

Equation~\eqref{eq:Cell} translates the compact coefficient system into explicit algebraic constraints. The reduced branch is obtained in three stages. The heavy branch is obtained from
\begin{equation}
0=\partial_XV=C_1(s)+2C_2(s)X+3C_3(s)X^2+\cdots .
\label{eq:Xsolvefull}
\end{equation}
When Eqs.~\eqref{eq:harmless1} and \eqref{eq:harmless2} hold, the leading solution is
\begin{equation}
X_v(s)=-{\frac{C_1(s)}{2C_2(s)}}
=-{\frac{B_{p+\nu}}{2A_p}}s^{-\nu}+\cdots .
\label{eq:XvC}
\end{equation}
Second, the reduced potential is
\begin{align}
U(s)&=V(X_v(s),s)\nonumber\\
&=C_0(s)-{\frac{C_1(s)^2}{4C_2(s)}}+
\Order\!\left({\frac{C_1^3C_3}{C_2^3}},C_3X_v^3,C_4X_v^4\right).
\label{eq:UfromC}
\end{align}
The negative term \(-C_1^2/(4C_2)\) is the energy released when the heavy coordinate relaxes to the stationary value. The ratio \(C_1/C_2\) determines the valley displacement, while \(C_1^2/C_2\) determines the energetic imprint of the displacement. A branch can rotate the trajectory appreciably even when the branch contribution to the plateau energy is subleading.
Third, the coefficients in
\begin{equation}
{\frac{U(s)}{V_0}}=1-{\frac{c_1}{s}}+{\frac{c_2}{s^2}}-{\frac{c_4^{\rm net}}{s^4}}-{\frac{c_8}{s^8}}+\cdots
\label{eq:Ubranch}
\end{equation}
are read from the large-$s$ expansion of Eq.~\eqref{eq:UfromC}. The microscopic matching conditions are
\begin{align}
 c_1&=-{\frac{1}{V_0}}[s^{-1}]\,U(s),\label{eq:c1match}\\
 c_2&={\frac{1}{V_0}}[s^{-2}]\,U(s),\label{eq:c2match}\\
 c_4^{\rm net}&=-{\frac{1}{V_0}}[s^{-4}]\,U(s),\label{eq:c4match}\\
 c_8&=-{\frac{1}{V_0}}[s^{-8}]\,U_{\rm exit}(s).\label{eq:c8match}
\end{align}
The notation \([s^{-n}]\) denotes the coefficient of \(s^{-n}\). The net coefficient $c_4^{\rm net}$ combines the $s^{-4}$ term in $C_0$ with the relaxation term in Eq.~\eqref{eq:UfromC}. Holding $N_*=55$, we propagate the official ACT tensor posterior through Eqs.~\eqref{eq:Ubranch} and \eqref{eq:Nintd} and apply the correlated running weight. The resulting marginal medians are $c_1=1.8432$, $c_2=3.5257$, and $c_8=6.6603\times10^7$. At the recalculated pivot, $|(c_8/s_*^8)/(c_1/s_*)|=1.53\times10^{-6}$. The exit term is therefore negligible for CMB curvature and becomes relevant near $s_{\rm end}$. Mirror quintic geometry determines the period basis. The integral representative in Eq.~\eqref{eq:fluxchoice} determines the bounded flux polynomial and the axiodilaton value entering Eqs.~\eqref{eq:Aj} through \eqref{eq:Cell}. The coefficients $(A_p,B_{p+\nu})$ follow from $X_v(s)$ and the normal mass in Eq.~\eqref{eq:mxscale}.

\section{Compact branch condition}

The period system, flux polynomial, and branch criterion define a finite arithmetic test.

\emph{Compact branch criterion.} Consider a Calabi Yau orientifold with one branch coordinate $X=e-ma$, odd flux lattice $\Lambda_{\rm odd}$, tadpole bound $L$, period vector $\Pi$, K\"ahler sector $K_K,W_K$ and stabilized spectators. If an integral flux pair \((f,h)\in\Lambda_{\rm odd}\) satisfies
\begin{align}
f^T\Sigma h+N_{D3}&=L,\label{eq:cond1}\\
C_0(s)&=V_0-c_qV_0s^{-q}+\cdots,\quad V_0>0,\quad c_q>0,\label{eq:cond2}\\
C_1(s)&=B_{p+\nu}s^{-(p+\nu)}+\cdots,\nonumber\\
C_2(s)&=A_ps^{-p}+\cdots,\quad A_p>0,\label{eq:cond3}\\
p+2\nu&>q,\qquad p<2,\nonumber\\
&\hbox{or}\quad p=2\quad\hbox{with}\quad {\frac{12A_pm^2}{dV_0}}\gg1.\label{eq:cond4}
\end{align}
and all higher operators obey Eqs.~\eqref{eq:harmless1} and \eqref{eq:harmless2}, then the branch admits a controlled single clock plateau with the large-$N_*$ predictions in Eqs.~\eqref{eq:nslarge} through \eqref{eq:alphalarge}. If the K\"ahler and tower data satisfy
\begin{equation}
m_H^2/H^2\gg1,
\qquad
V_{\rm bar}/V_*>1,
\qquad
\Lambda_{\rm EFT}/H\gg1,
\label{eq:cond5}
\end{equation}
the reduced branch is stable under heavy sector integration.

The criterion compares the inflationary energy with the heavy sector mass gap and the EFT cutoff. Eqs.~\eqref{eq:cond2} through \eqref{eq:cond4} encode the displaced valley and entropy mass conditions. The higher operator inequalities preserve the leading exponents. The tadpole equation determines the available flux charge, while Eq.~\eqref{eq:cond5} bounds the Schur correction induced by the integrated heavy fields. Compactification data enter through the finite Diophantine system $C_\ell(s,f,h,S)$ on $\Lambda_{\rm odd}$. A viable branch must satisfy the tadpole equation, the coefficient map, and the heavy sector inequalities simultaneously. Within the stated hierarchy, these conditions are sufficient for the reduced branch. Each integral solution maps \((f,h,S,\Pi)\) to a definite coefficient vector.

Flux vacuum counting and tadpole cancellation restrict the integral lattice before the scalar potential is minimized~\cite{AshokDouglas:2004,DenefDouglas:2004}. Explicit Calabi Yau flux compactifications exhibit the corresponding arithmetic relation between fluxes, moduli, and D3 charge~\cite{Giryavets:2004}. The active pair in Eq.~\eqref{eq:Nfluxzero} has $N_{\rm flux}=0$. The D3 charge budget can therefore be assigned to spectator stabilization, localized sources, and mobile D3 branes. Tadpole pressure enters through that spectator sector~\cite{Bena:2018}. A global odd lattice must reproduce the coefficient image, tadpole balance, and spectator mass gap in Eq.~\eqref{eq:cond5}.

\section{K\"ahler volume stabilization}

\subsection{Volume dressing}

The four dimensional $\mathcal N=1$ F term potential has the standard supergravity form~\cite{GKP:2002,KKLT:2003,DouglasKachru:2007}
\begin{equation}
V=e^K\left(K^{A\bar B}D_AW D_{\bar B}\bar W-3|W|^2\right)+V_{\rm up},
\label{eq:SUGRAV}
\end{equation}
with $A,B\in\{z,S,T\}$ in the one active complex structure truncation. The K\"ahler potential and superpotential are
\begin{align}
K&=K_{\cs}(z,\bar z)+K_S(S,\bar S)+K_K(T,\bar T),\label{eq:Ksplit}\\
W&=W_{\rm flux}(z,S)+A_{\rm np}e^{-aT}.
\label{eq:Wsplit}
\end{align}
The F term contribution is
\begin{align}
V_F=e^K\Big(&K^{z\bar z}|D_zW|^2+K^{S\bar S}|D_SW|^2\nonumber\\
&+K^{T\bar T}|D_TW|^2-3|W|^2\Big).
\label{eq:VFcomplete}
\end{align}
Every covariant derivative in Eq.~\eqref{eq:VFcomplete} contains the total superpotential. Flux and nonperturbative sectors interact through both \(D_AW\) and the common supergravity factor \(e^K\). The heavy field reduction begins from Eq.~\eqref{eq:VFcomplete}. Two independent additive scalar potentials omit the mixed response responsible for the volume shift and the light heavy Hessian entries.

Let $S_v(s)$ and $T_v(s)$ denote the adiabatic heavy solution at a specified branch coordinate. Expanding Eq.~\eqref{eq:VFcomplete} around the adiabatic solution separates the stabilized K\"ahler potential from the active complex structure energy at leading order in $H^2/m_H^2$. The complex structure F terms inherit the factor $e^{K_K}=Y^{-2}$, so a branch energy normalized at $\tau_0$ is
\begin{equation}
V_{\rm br}(s,\tau)=V_0F(s){\frac{Y(\tau_0)^2}{Y(\tau)^2}}.
\label{eq:Vbrdress}
\end{equation}
Equation~\eqref{eq:Vbrdress} is the leading term of the adiabatic expansion of the combined potential. The normalization follows from the common supergravity potential. The factor \(Y^{-2}\) has a direct dynamical meaning. Inflationary energy exerts a force toward larger volume because increasing \(Y\) lowers the branch energy. The K\"ahler curvature opposes the volume force and determines the displacement \(\Delta\tau\). The mixed Hessian and Schur complement quantify the residual feedback on the adiabatic curvature.

The reduced uplift term is parameterized as
\begin{equation}
V_{\rm up}={\frac{D}{(2\tau)^2}},
\label{eq:Vup}
\end{equation}
consistent with warped and D term uplift scalings in four dimensional effective descriptions~\cite{KPV:2002,Burgess:2003}. The coefficient $D$ is determined by $V_K(\tau_0,0)=0$ and $\partial_\tau V_K(\tau_0,0)=0$ for the chosen nonperturbative sector.

\subsection{BBHL K\"ahler geometry}

The mirror quintic has $h^{1,1}=101$~\cite{Candelas:1991}. The field $T$ below denotes the overall volume direction after the orthogonal K\"ahler directions are integrated out with positive masses. The reduced K\"ahler block describes the response of the overall volume. The separation between the overall volume and orthogonal K\"ahler directions follows the hierarchy used in large volume stabilization analyses~\cite{LVS:2005,CicoliConlonQuevedo:2008}.

Let
\begin{equation}
T=\tau+i\rho,
\qquad
\cV=\tau^{3/2},
\qquad
Y(\tau)=\cV+{\frac{\hat\xi}{2}}.
\end{equation}
The Becker Becker Haack Louis correction gives the leading $\alpha'^3$ K\"ahler potential~\cite{BBHL:2002}
\begin{equation}
K_K=-2\log Y(\tau),
\qquad
\hat\xi={\frac{-\chi(Y)\zeta(3)}{2(2\pi)^3g_s^{3/2}}}.
\label{eq:BBHLK}
\end{equation}
The derivatives entering the F term potential are
\begin{align}
K_T&={\frac{1}{2}}\partial_\tau K_K=-{\frac{3\sqrt\tau}{2Y}},\label{eq:KT}\\
K_{T\bar T}&={\frac{1}{4}}\partial_\tau^2K_K=-{\frac{3}{8\sqrt\tau Y}}+{\frac{9\tau}{8Y^2}},\label{eq:KTT}\\
K^{T\bar T}&=(K_{T\bar T})^{-1}.\label{eq:Kinv}
\end{align}
At a specified $s$ on the heavy relaxation time scale, $W_0$ denotes the constant flux contribution to the K\"ahler block after the heavy flux fields reach their stationary values. It is distinct from the active polynomial value $W_*$ in Eq.~\eqref{eq:activeDtW}. The nonperturbative superpotential follows the KKLT form~\cite{KKLT:2003}. On the real axion plane,
\begin{equation}
W_{\rm eff}=W_0+A_{\rm np}e^{-a\tau}e^{-ia\rho}.
\label{eq:WKrho}
\end{equation}
The K\"ahler covariant derivative of the superpotential is
\begin{equation}
D_TW_{\rm eff}=K_TW_0+(-a+K_T)A_{\rm np}e^{-a\tau}e^{-ia\rho}.
\label{eq:DTWfull}
\end{equation}
Define
\begin{equation}
\mathcal A(\tau)=K_TW_0,
\qquad
\mathcal B(\tau)=(-a+K_T)A_{\rm np}e^{-a\tau}.
\end{equation}
Then
\begin{align}
|D_TW_{\rm eff}|^2&=\mathcal A^2+\mathcal B^2+2\mathcal A\mathcal B\cos(a\rho),\label{eq:DTWcos}\\
|W_{\rm eff}|^2&=W_0^2+A_{\rm np}^2e^{-2a\tau}+2A_{\rm np}W_0e^{-a\tau}\cos(a\rho).
\label{eq:Wcos}
\end{align}
The reduced K\"ahler block is
\begin{equation}
V_F^{(K)}=e^{K_K}\left(K^{T\bar T}|D_TW_{\rm eff}|^2-3|W_{\rm eff}|^2\right),
\label{eq:VKfull}
\end{equation}
and
\begin{equation}
V_K(\tau,\rho)=V_F^{(K)}(\tau,\rho)+{\frac{D}{(2\tau)^2}}.
\label{eq:VKup}
\end{equation}
The trigonometric structure of Eqs.~\eqref{eq:DTWcos} and \eqref{eq:Wcos} enforces $\rho=0$ as an invariant axis. Every derivative linear in the $\rho$ displacement is proportional to $\sin(a\rho)$, so
\begin{equation}
\partial_\rho V_K|_{\rho=0}=0,
\qquad
\partial_\tau\partial_\rho V_K|_{\rho=0}=0.
\label{eq:mixedzero}
\end{equation}
The real metric is
\begin{equation}
G_{\tau\tau}=G_{\rho\rho}=2K_{T\bar T},
\end{equation}
and the canonical mass matrix is
\begin{equation}
(M_K^2)^i{}_{j}=G^{ik}\nabla_k\nabla_jV_K,
\qquad i,j\in\{\tau,\rho\}.
\label{eq:KMmatrix}
\end{equation}
Because the background solves $\partial_iV_K=0$, the Christoffel connection drops out of the covariant Hessian at the minimum, so the canonical mass matrix measures the microscopic curvature of the K\"ahler potential.

The benchmark values
\begin{align}
\tau_0&=100, & a&=0.08, & A_{\rm np}&=10,\nonumber\\
W_0&=-2.32741322\times10^{-2},\nonumber\\
D&=4.80522792\times10^{-5}
\label{eq:Kparams}
\end{align}
solve
\begin{equation}
V_K(\tau_0,0)=0,
\qquad
\partial_\tau V_K(\tau_0,0)=0,
\qquad
\partial_\rho V_K(\tau_0,0)=0.
\end{equation}
For \(\chi(Y)=+200\) and \(g_s=0.079861\),
\begin{equation}
{\frac{\hat\xi}{2\cV(\tau_0)}}=-1.074\times10^{-2},
\label{eq:xirat}
\end{equation}
so the BBHL correction changes the logarithmic volume geometry at the percent level. The negative sign follows from the positive Euler characteristic of the mirror quintic. Retuning $W_0$ and $D$ in Eq.~\eqref{eq:Kparams} preserves the Minkowski stationary point and the volume barrier with the corrected BBHL geometry.

\subsection{Barrier and decoupling}

The total potential used for the inflationary branch is
\begin{equation}
V_{\rm tot}(s,\tau,\rho)=V_K(\tau,\rho)+V_0F(s){\frac{Y(\tau_0)^2}{Y(\tau)^2}},
\label{eq:VtotK}
\end{equation}
with $F(s)=U(s)/V_0$. Expanding the $\tau$ equation around the stabilized point yields the leading displacement of the volume modulus.
\begin{equation}
0=\partial_\tau V_{\rm tot}\simeq \partial_\tau^2V_K|_0\,\Delta\tau+
\partial_\tau\!\bigg[V_0F(s){\frac{Y(\tau_0)^2}{Y(\tau)^2}}\bigg]_{\tau_0},
\end{equation}
so
\begin{equation}
\Delta\tau(s)=-{\frac{\partial_\tau[V_0F(s)Y(\tau_0)^2/Y(\tau)^2]_{\tau_0}}{\partial_\tau^2V_K(\tau_0,0)}}+
\Order\left({\frac{H^4}{m_\tau^4}}\right).
\label{eq:dtau}
\end{equation}
The volume response has two independent control conditions. The local condition bounds the displacement around \(\tau_0\). The global condition compares the inflationary energy with the neighboring decompactification saddle,
\begin{equation}
\partial_\tau V_K(\tau_{\rm max},0)=0,
\qquad
V_{\rm bar}=V_K(\tau_{\rm max},0)-V_K(\tau_0,0).
\label{eq:barrier}
\end{equation}
The benchmark ratio \(V_{\rm bar}/V_*=7.202\) at \(N_*=55\) leaves the full trajectory below the escape saddle. The displacement \(\Delta\tau\) measures local volume response, and the barrier measures resistance to decompactification. The benchmark satisfies both independent requirements.

The axiodilaton value of the integral representative follows from $D_SW=0$ at the benchmark saxion, as shown in Eq.~\eqref{eq:Sflux}. The covariant flux sector Hessian in the stationary theory is
\begin{equation}
(M_{\rm flux}^2)^A{}_{B}=G^{A\bar C}\nabla_{\bar C}\nabla_BV,
\qquad A,B\in\{S,z\}.
\label{eq:fluxmass}
\end{equation}
The explicit stability calculation below evaluates the branch normal direction and the overall volume pair. The common positive mass requirement for additional spectator directions is Eq.~\eqref{eq:cond3}, while $M_{\rm KK}$ and $M_s$ test the separation from the internal compactification scales.

Let the light variable be $\ell$ and the heavy block be $H=(X,\tau,\rho,S,\cs_\perp)$. The covariant mass matrix has block form
\begin{equation}
\mathcal M^2=\begin{pmatrix}M^2_{\ell\ell}&M^2_{\ell H}\\ M^2_{H\ell}&M^2_{HH}\end{pmatrix}.
\end{equation}
Integrating out the heavy block produces the Schur complement
\begin{equation}
M^2_{\rm eff}=M^2_{\ell\ell}-M^2_{\ell H}(M^2_{HH})^{-1}M^2_{H\ell}.
\label{eq:Schur}
\end{equation}
For a positive heavy block, the second term is negative semidefinite. Heavy relaxation lowers or preserves the light curvature at quadratic order. The smallest heavy eigenvalue \(\mu_H\) and the mixing norm \(\|M^2_{\ell H}\|\) obey
\begin{equation}
\left|M^2_{\rm eff}-M^2_{\ell\ell}\right|\le {\frac{\|M^2_{\ell H}\|^2}{\mu_H}}.
\label{eq:SchurBound}
\end{equation}
For the benchmark the lowest integrated threshold is the branch mass, \(m_X^2/H^2\simeq35\), while the K\"ahler eigenvalues exceed \(3\times10^3\). Differentiating the volume dressed branch energy in Eq.~\eqref{eq:Vbrdress} determines the canonical light volume mixing. Table~\ref{tab:schur} reports an induced shift in \(\eta\) of order \(10^{-6}\) throughout the CMB interval. The resulting hierarchy is quantitative. The \(s^{-2}\) flux curvature changes \(n_s\) at the retained precision, while K\"ahler relaxation changes \(n_s\) more than four orders of magnitude less.

\begin{table}[t]
\caption{Schur complement decoupling check for the K\"ahler block. The mixing entry is the canonical light volume Hessian in Hubble units. The last column includes the factor \(H^2/V\simeq1/3\) converting the Schur correction to the slow roll curvature $\eta$.}
\label{tab:schur}
\begin{ruledtabular}
\begin{tabular}{cccc}
$N_*$ & $|M^2_{\varphi\tau}|/H^2$ & $m_\tau^2/H^2$ & $|\Delta\eta_{\rm Schur}|$ \\
\hline
50 & $1.42\times10^{-1}$ & $3.02\times10^3$ & $2.23\times10^{-6}$ \\
55 & $1.32\times10^{-1}$ & $3.51\times10^3$ & $1.66\times10^{-6}$ \\
60 & $1.23\times10^{-1}$ & $4.03\times10^3$ & $1.26\times10^{-6}$ \\
65 & $1.16\times10^{-1}$ & $4.59\times10^3$ & $9.70\times10^{-7}$ \\
\end{tabular}
\end{ruledtabular}
\end{table}

\begin{table*}[t]
\caption{Stability data for the BBHL corrected K\"ahler sector, the heavy branch, and the compactification thresholds. The K\"ahler eigenvalues and branch mass are reported in Hubble units. The KK and string entries use the Einstein frame normalization in Eq.~\eqref{eq:MsKK}.}
\label{tab:stability}
\begin{ruledtabular}
\begin{tabular}{cccccccc}
$N_*$ & $m_\tau^2/H^2$ & $m_\rho^2/H^2$ & $m_X^2/H^2$ & $M_{\rm KK}/H$ & $M_s/H$ & $|\Delta\tau|/\tau_0$ & $V_{\rm bar}/V_*$ \\
\hline
50 & $3.02\times10^3$ & $3.34\times10^3$ & $35.0$ & $2.68\times10^2$ & $8.49\times10^2$ & $1.95\times10^{-3}$ & $6.21$ \\
55 & $3.51\times10^3$ & $3.87\times10^3$ & $35.0$ & $2.89\times10^2$ & $9.14\times10^2$ & $1.68\times10^{-3}$ & $7.20$ \\
60 & $4.03\times10^3$ & $4.45\times10^3$ & $35.0$ & $3.10\times10^2$ & $9.79\times10^2$ & $1.47\times10^{-3}$ & $8.27$ \\
65 & $4.59\times10^3$ & $5.06\times10^3$ & $35.0$ & $3.30\times10^2$ & $1.05\times10^3$ & $1.29\times10^{-3}$ & $9.42$ \\
\end{tabular}
\end{ruledtabular}
\end{table*}

\begin{figure*}[!t]
\includegraphics[width=0.96\textwidth]{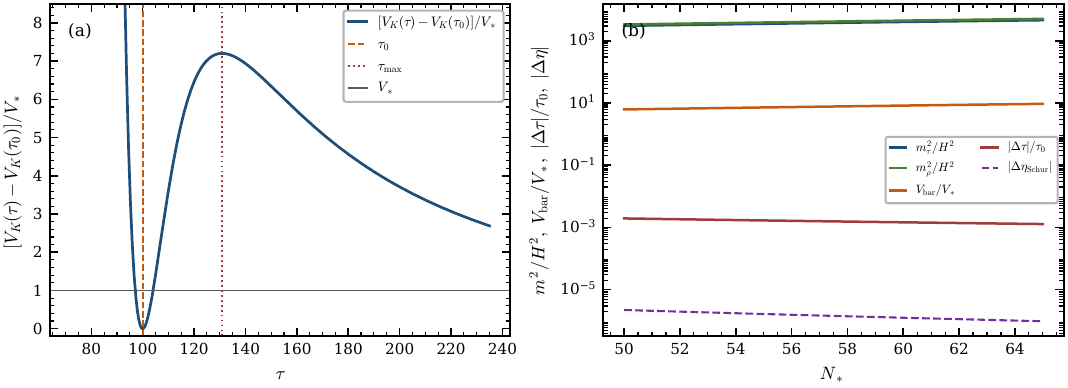}
\caption{K\"ahler barrier and heavy sector decoupling. The left panel displays the uplifted BBHL corrected KKLT potential around the stabilized volume in units of $V_*$. The right panel displays the K\"ahler masses, fractional volume displacement, barrier ratio, and Schur correction across the CMB interval.}
\label{fig:kahlerbarrier}
\end{figure*}

\section{Reduced branch and ACT posterior scan}

\begin{figure*}[!t]
\includegraphics[width=0.96\textwidth]{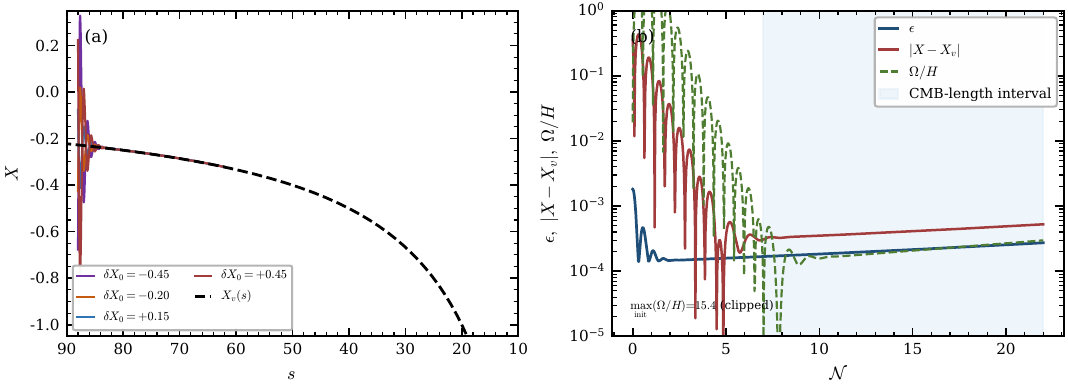}
\caption{Field space trajectories and CMB window evolution. The left panel displays covariant two field solutions in the $(s,X)$ plane from several initial heavy displacements. The right panel displays the slow roll parameters and orthogonal displacement histories, confirming relaxation before the CMB observables are evaluated.}
\label{fig:fieldevol}

\end{figure*}

The mirror quintic branch used for the benchmark is
\begin{align}
{\frac{U(s)}{V_0}}&=1-{\frac{1.8432}{s}}+{\frac{3.5257}{s^2}}-{\frac{c_4^{\rm net}}{s^4}}-{\frac{6.6603\times10^7}{s^8}},\nonumber\\
c_4^{\rm net}&=0.
\label{eq:branchd3}
\end{align}
The central coefficients \(1.8432\), \(3.5257\), and \(6.6603\times10^7\) are the marginal medians obtained by conditioning the official ACT posterior propagation on $N_*=55$, as described in Sec.~\ref{sec:ACTprop}. They satisfy the weight conditions in Eq.~\eqref{eq:weightthm} and the exit matching definition in Eq.~\eqref{eq:c8match}. The numerical coefficients acquire physical meaning when multiplied by the associated saxion powers. The \(s^{-1}\) term governs the leading approach to the plateau. The \(s^{-2}\) term deforms the canonical curvature entering \(\eta\). The \(s^{-8}\) term regulates the exit and remains negligible at the pivot. Because \(\epsilon\) depends on the squared gradient and \(\eta\) depends linearly on curvature, the \(s^{-2}\) operator raises the scalar tilt with a much smaller change in the tensor amplitude. Growth of the \(s^{-8}\) term steepens the gradient after the CMB modes have exited. Covariant integrations in Figure~\ref{fig:fieldevol} show continued alignment of the heavy coordinate with the stationary branch throughout the observable interval. At the pivot,
\begin{align}
s_*&=81.39,\nonumber\\
\left|{\frac{(3.5257/s_*^2)}{(1.8432/s_*)}}\right|&=2.35\times10^{-2},\nonumber\\
\left|{\frac{(6.6603\times10^7/s_*^8)}{(1.8432/s_*)}}\right|&=1.53\times10^{-6}.
\end{align}
Equation~\eqref{eq:Us} determines the branch relaxation weight at \(s^{-(p+2\nu)}=s^{-4}\). For the stabilizing pair in Eq.~\eqref{eq:Vbr3D}, $A_2/V_0=8.75$ and $\beta=B_3/(2A_2)=20$, so
\begin{align}
c_4^{\rm br}&={\frac{B_3^2}{4A_2V_0}}={\frac{A_2}{V_0}}\beta^2\nonumber\\
&={\frac{d}{12}}{\frac{m_X^2}{H^2}}\beta^2=3500,\nonumber\\
c_4^{\rm net}&=c_4^{\rm br}-c_4^{\rm comp}.
\label{eq:c4consistency}
\end{align}
Expanding the positive square in Eq.~\eqref{eq:Vbr3D} contributes $c_4^{\rm comp}=3500$ to $C_0$, while stationary relaxation subtracts $c_4^{\rm br}=3500$. The central square construction therefore has $c_4^{\rm net}=0$. Allowing $0\le c_4^{\rm net}\le3500$ defines the correlated relaxation envelope. The upper endpoint contributes $3.52\times10^{-3}$ relative to the leading falloff when evaluated at the central pivot and $2.54\times10^{-3}$ at its self consistent pivot. At $N_*=55$ it gives $s_{\rm end}=13.1063$, $s_*=90.7309$, $n_s=0.972762$, $r=2.14733\times10^{-3}$, and $\alpha_s=-3.96462\times10^{-4}$. Relative to the central completion, the shifts are $\Delta n_s=1.71\times10^{-3}$, $\Delta r/r=-17.34\%$, and $\Delta\alpha_s=9.78\times10^{-6}$. Equation~\eqref{eq:c4consistency} varies the residual jointly with the displacement and normal mass. The numerical ratios separate the three contributions. The curvature operator is visible during the CMB interval, while the exit operator is six orders smaller than the leading falloff at the pivot. Near \(s_{\rm end}\), the high power compensates for the large coefficient and steepens the gradient. The high power produces the exit after horizon crossing has measured the plateau curvature.

\begin{figure*}[!t]
\includegraphics[width=0.96\textwidth]{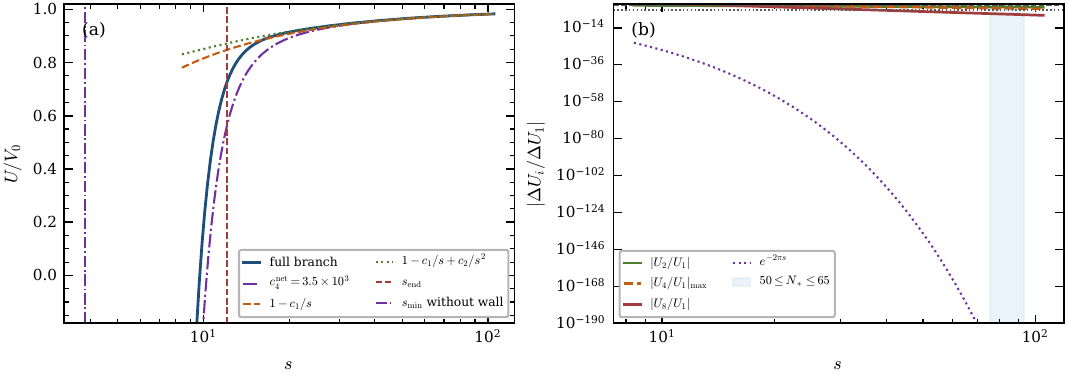}
\caption{Reduced branch and expansion ordering. The left panel displays $\mathcal U/V_0$ for the central square branch and the upper residual $c_4^{\rm net}=3500$ envelope, together with successive inverse saxion truncations. The right panel displays the $s^{-2}$ curvature, maximum residual $s^{-4}$ relaxation, and $s^{-8}$ exit contributions relative to the leading $s^{-1}$ term over the CMB interval.}
\label{fig:expansion}
\end{figure*}

The scalar spectral index, tensor to scalar ratio, and running are defined by
\begin{equation}
n_s-1={\frac{\dd\log\mathcal P_{\cal R}}{\dd\log k}},
\qquad
r={\frac{\mathcal P_T}{\mathcal P_{\cal R}}},
\qquad
\alpha_s={\frac{\dd n_s}{\dd\log k}}.
\label{eq:observabledefs}
\end{equation}
The three observables probe different derivatives of the background~\cite{Dodelson:2003}. The ratio \(r\) measures the geodesic slope through \(\epsilon\). The tilt \(n_s\) also measures the geodesic curvature through \(\eta\). The running \(\alpha_s\) measures variation of the geodesic curvature across neighboring horizon exit times. The derivative hierarchy explains why a subleading \(s^{-2}\) term can shift \(n_s\) more strongly than \(r\).

For general \(d\),
\begin{align}
U_{,\varphi}&=\sqrt{\frac{2}{d}}sU_{,s},\label{eq:Ud1}\\
U_{,\varphi\varphi}&={\frac{2}{d}}s(U_{,s}+sU_{,ss}),\label{eq:Ud2}\\
U_{,\varphi\varphi\varphi}&=\sqrt{\frac{2}{d}}s{\frac{\dd}{\dd s}}
\left[{\frac{2}{d}}s(U_{,s}+sU_{,ss})\right].\label{eq:Ud3}
\end{align}
The conversion from \(s\) derivatives to canonical derivatives inserts one factor \(\sqrt{2/d}\) for every derivative along the logarithmic geodesic. The spectral formulas then follow from the standard slow roll expansion~\cite{Dodelson:2003,LiddleLyth:2000}. The Hodge degree enters the observables through the kinetic geometry, and a rescaling of \(U\) leaves the conversion factor unchanged.
The standard slow roll observables then read~\cite{LiddleLyth:2000,Dodelson:2003}
\begin{align}
\epsilon&={\frac{1}{2}}\left({\frac{U_{,\varphi}}{U}}\right)^2,
&r&=16\epsilon,\label{eq:epsr}\\
\eta&={\frac{U_{,\varphi\varphi}}{U}},
&n_s&=1-6\epsilon+2\eta,\label{eq:etans}\\
\xi_2&={\frac{U_{,\varphi}U_{,\varphi\varphi\varphi}}{U^2}},
&\alpha_s&=16\epsilon\eta-24\epsilon^2-2\xi_2.\label{eq:xis}
\end{align}
The gradient enters \(r\) quadratically, while the curvature enters \(n_s\) linearly. A moderate curvature deformation can move the scalar tilt across an observational contour with little change in the tensor amplitude. The third derivative enters through \(\xi_2\), explaining the small running of a smoothly deformed plateau.

The e fold integral is
\begin{equation}
N_*={\frac{d}{2}}\int_{s_{\rm end}}^{s_*}{\frac{U}{s^2U_{,s}}}\,\dd s.
\label{eq:Nintd}
\end{equation}
The factor \(d/2\) converts saxion evolution into canonical e fold time. The Hodge degree enters both the field distance and the gradient normalization. It changes the field distance accumulated between \(s_{\rm end}\) and \(s_*\), and the degree changes the gradient normalization entering \(r\). The pivot location and tensor amplitude are two manifestations of one boundary metric. At \(N_*=55\), evaluating Eq.~\eqref{eq:branchd3} yields
\begin{align}
n_s&=0.97105, & r&=2.60\times10^{-3},\nonumber\\
\alpha_s&=-4.06\times10^{-4}.
\label{eq:d3obs}
\end{align}
The analytic $d=1$ reference uses $(c_1,c_2,c_8)=(10,580,10^{18})$. The coefficients normalize a 55 e fold trajectory with a controlled exit, so the comparison isolates the observable change induced by the canonical factor $d$. The analytic reference yields
\begin{align}
n_s&=0.97468, & r&=7.65\times10^{-4},\nonumber\\
\alpha_s&=-2.85\times10^{-4}.
\label{eq:d1obs}
\end{align}
using the $d=1$ canonical normalization. The mirror quintic coefficient map applies to the $d=3$ branch. Figure~\ref{fig:nsr} displays the two geometric classes separately.

Table~\ref{tab:componentroles} separates the curvature deformation from the exit wall at $N_*=55$. With $c_8=0$ and $c_2>0$, the reduced potential approaches a stationary minimum at $s=2c_2/c_1=3.83$ and approaches the minimum before $\epsilon$ reaches unity. The wall produces the physical exit. Once the endpoint has been determined, the positive $s^{-2}$ coefficient changes the scalar curvature while leaving the tensor scale at order $10^{-3}$.

\begin{table}[t]
\caption{Component separation in the reduced \(d=3\) branch at \(N_*=55\). The table isolates the \(s^{-2}\) curvature deformation, the residual \(s^{-4}\) relaxation envelope, and the \(s^{-8}\) exit wall in Eq.~\eqref{eq:branchd3}. The wall determines the endpoint and the \(N_*\)-to-\(s_*\) map, while the curvature and residual relaxation terms produce the controlled finite distance shifts.}
\label{tab:componentroles}
\scriptsize
\begin{ruledtabular}
\begin{tabular}{ccccc}
$(c_2,c_4^{\rm net},c_8)$ & $s_{\rm end}$ & $s_*$ & $n_s$ & $r$ \\
\hline
$(0,0,0)$ & $2.91$ & $76.52$ & $0.96587$ & $3.25\times10^{-3}$ \\
$(3.5257,0,0)$ & \multicolumn{4}{c}{minimum at $s=3.83$, no exit} \\
$(0,0,6.6603\times10^7)$ & $12.20$ & $87.66$ & $0.97044$ & $2.46\times10^{-3}$ \\
$(3.5257,0,6.6603\times10^7)$ & $12.11$ & $81.39$ & $0.97105$ & $2.60\times10^{-3}$ \\
$(3.5257,3500,6.6603\times10^7)$ & $13.11$ & $90.73$ & $0.97276$ & $2.15\times10^{-3}$ \\
\end{tabular}
\end{ruledtabular}
\end{table}

Figure~\ref{fig:fieldevol} extends the branch comparison in Figures~\ref{fig:branch_diag} and \ref{fig:lw3dcomparison} to covariant two field evolution of the reduced \(d=3\) branch. Initial orthogonal displacements relax toward \(X_v(s)\) before horizon exit across \(50\le N_*\le65\), and the slow roll parameters remain small throughout the observable interval. The trajectory follows the logarithmic saxionic attractor while the invariant coordinate is the massive transverse response.

\subsection{Invariant attractor equation}

\begin{figure*}[!t]
\includegraphics[width=0.96\textwidth]{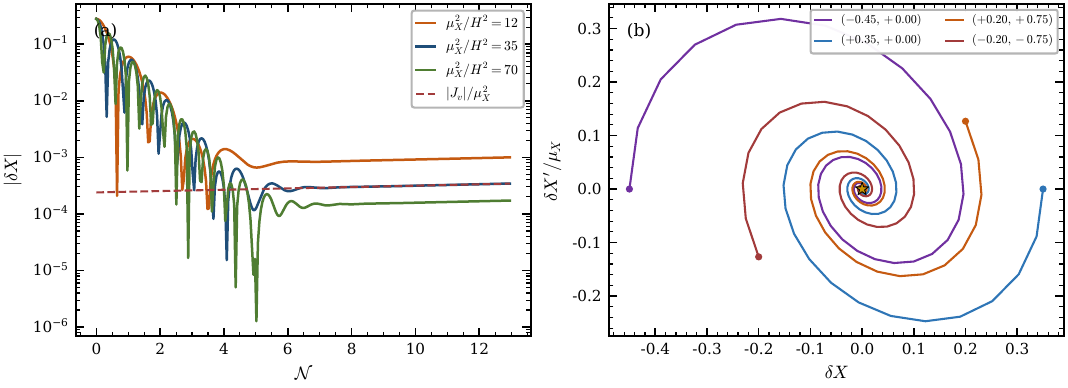}
\caption{Invariant attractor equation for the heavy branch coordinate. The left panel displays solutions of Eq.~\eqref{eq:invariantAttractor} for several $\mu_X^2/H^2$ together with the source $J_v$. The right panel displays phase space relaxation in the $(s,X)$ plane for different initial displacements and velocities.}
\label{fig:invariantattractor}
\end{figure*}

The dynamics of the displaced valley are most transparently formulated in the monodromy invariant variable. For the metric in Eq.~\eqref{eq:metricXs}, the $a$ equation in e fold time ${\cal N}=\log a_{\rm FRW}$ is
\begin{equation}
a''+\left(3-\epsilon-2{\frac{s'}{s}}\right)a'
+{\frac{2s^2}{dH^2}}\,\partial_aV=0,
\label{eq:aEOM}
\end{equation}
A prime denotes \(\dd/\dd{\cal N}\). Multiplication by $-m$ and the identity $X=e-ma$ yield
\begin{equation}
X''+\left(3-\epsilon-2{\frac{s'}{s}}\right)X'
+{\frac{2m^2s^2}{dH^2}}\,\partial_XV=0 .
\label{eq:Xgeneral}
\end{equation}
For the leading displaced potential,
\begin{equation}
V\supset A s^{-2}X^2+B s^{-3}X,
\qquad
X_v(s)=-{\frac{B}{2A}}{\frac{1}{s}},
\end{equation}
Eq.~\eqref{eq:Xgeneral} reduces to
\begin{align}
X''+\left(3-\epsilon-2{\frac{s'}{s}}\right)X'
+\mu_X^2\,[X-X_v(s)]&=0,
\label{eq:invariantAttractor}\\
\mu_X^2&={\frac{4Am^2}{dH^2}} .
\end{align}
Writing $\delta X=X-X_v(s)$ casts the equation into the driven form
\begin{equation}
\delta X''+\left(3-\epsilon-2{\frac{s'}{s}}\right)\delta X'
+\mu_X^2\delta X=-J_v,
\label{eq:deltaXeq}
\end{equation}
with
\begin{equation}
J_v=X_v''+\left(3-\epsilon-2{\frac{s'}{s}}\right)X_v' .
\end{equation}
Equation~\eqref{eq:deltaXeq} has the form of a damped oscillator with a moving equilibrium point. The source \(J_v\) measures the acceleration of the stationary valley in e fold time, while \(\mu_X^2\) measures the restoring response normal to the valley. In the adiabatic regime,
\begin{equation}
\delta X_{\rm lag}\simeq-{\frac{J_v}{\mu_X^2}},
\label{eq:adiabaticlag}
\end{equation}
so the heavy mass suppresses dependence on the initial displacement and reduces the lag generated by valley motion. The benchmark value \(\mu_X^2\simeq35\) exceeds both Hubble friction and the driving rate. Figure~\ref{fig:invariantattractor} integrates Eq.~\eqref{eq:invariantAttractor} along the \(d=3\) background for several masses, initial displacements, and initial heavy velocities. Decay of \(|X-X_v|\) and the phase space spirals demonstrate attraction before the CMB observables are evaluated.

\begin{figure*}[!t]
\includegraphics[width=0.96\textwidth]{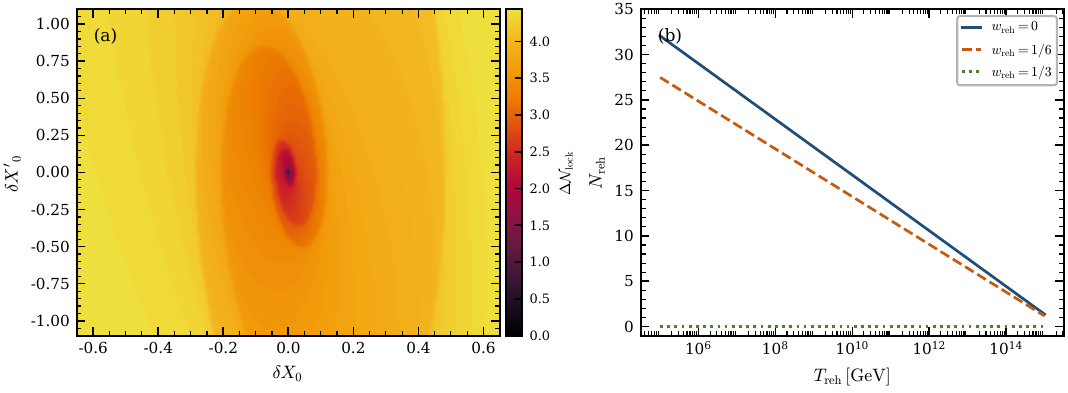}
\caption{Basin attraction and reheating duration derived from the background equations. The left panel displays the locking time $\Delta\mathcal N$ over the initial heavy phase space at the increased numerical resolution. The right panel displays $N_{\rm reh}$ from Eq.~\eqref{eq:reheat}. Radiation like reheating with $w_{\rm reh}=1/3$, gives $N_{\rm reh}=0$ and is shown as the horizontal line.}
\label{fig:basinreheating}
\end{figure*}

The scan uses $d=3$ and samples
\begin{align}
c_1&\in[1.0,2.4], & c_2&\in[-5,20],\nonumber\\
\log_{10}c_8&\in[7.2,8.4].
\label{eq:scanranges}
\end{align}
The scan retains points satisfying
\begin{equation}
\left|{\frac{c_2/s_*^2}{c_1/s_*}}\right|<0.25,
\qquad
\left|{\frac{c_8/s_*^8}{c_1/s_*}}\right|<5\times10^{-4}.
\label{eq:cuts}
\end{equation}
Positive $c_2$ moves $n_s$ upward because
\begin{align}
\eta&={\frac{2}{d}}{\frac{s(U_s+sU_{ss})}{U}}\nonumber\\
&=-{\frac{2c_1}{ds}}+{\frac{8c_2-2c_1^2}{ds^2}}
+\Order(s^{-3},c_8s^{-8}).
\label{eq:etashift}
\end{align}
The combination \(8c_2-2c_1^2\) is the finite distance curvature correction. The positive value shifts \(\eta\) upward. The corresponding change in \(\epsilon\) enters at higher relative order because \(\epsilon\) squares the first derivative. The flux coefficient \(c_2\) displaces the ACT posterior ensemble mainly along the \(n_s\) direction, while the Hodge degree continues to control the tensor band.

The leading relation between \(n_s\) and \(r\) joins the attractor plateau class after the branch has been reduced along the saxionic direction. In alpha attractor models the exponential coefficient follows from the hyperbolic pole in the inflaton metric~\cite{Kallosh:2013,KalloshLindeRoest:2013}. For the displaced branch, the coefficient is the boundary degree \(d\) extracted from the Hodge norm. The relation establishes a microscopic attractor correspondence. Bottom up $\alpha$ attractor models treat the pole strength as a continuous supergravity parameter, whereas the displaced Calabi Yau branch determines $\alpha_{\rm pole}=d/(3q^2)$ through the Hodge degree and derives the subleading correction from the flux lattice. The retained \(s^{-2}\) term is the leading flux lattice fingerprint shifting the tilt away from the pure topological attractor~\cite{Roest:2015,MartinRingeval:2014}.

\section{Official ACT posterior propagation, running and reheating}\label{sec:ACTprop}

The Hodge degree $d=3$ determines the leading attractor baseline, while the reduced coefficients control the finite distance curvature. We therefore map the official ACT DR6.02 posterior products into the branch parameter space~\cite{ACTDR6LCDM,ACTDR6Extended,ACTDR6Chains}. The weighted $(n_s,r)$ density is reconstructed from the Planck and ACT lite tensor chain with BK18, CMB lensing, and DESI BAO. The weighted $(n_s,\alpha_s)$ density is reconstructed from the Planck and ACT lite NRUN chain. Holding $N_*=55$, the tensor product defines the Metropolis target for the deterministic map $(c_1,c_2,c_8)\mapsto(n_s,r,\alpha_s)$. The NRUN density contributes a correlated weight because the products share CMB likelihoods. Smooth Planck and ACT boundaries are retained for display, whereas all sampling weights are calculated from the official chains.

Conditioning on $N_*$, the calculation samples $(c_1,c_2,\log_{10}c_8)$ over Eq.~\eqref{eq:scanranges}, subject to Eq.~\eqref{eq:cuts}, with $c_4^{\rm net}=0$. Six chains retain $32\,400$ samples with mean acceptance $0.663$ and largest split $\hat R=1.002$. The componentwise marginal medians define the central benchmark,
\begin{equation}
(c_1,c_2,\log_{10}c_8)=(1.8432,3.5257,7.8235),
\label{eq:ACTbenchmarkcoeff}
\end{equation}
which gives
\begin{equation}
(n_s,r,\alpha_s)=(0.97105,2.60\times10^{-3},-4.06\times10^{-4}).
\label{eq:ACTbenchmarkobs}
\end{equation}
The recalculated point lies inside the $68.27\%$ highest density region of each official posterior product. A second calculation allows $50\le N_*\le65$ to propagate reheating uncertainty. Its $32\,400$ retained samples have mean acceptance $0.630$, largest split $\hat R=1.002$, and posterior medians
\begin{equation}
(c_1,c_2,\log_{10}c_8,N_*)=(1.92,3.54,7.81,58.00),
\end{equation}
with $(n_s,r,\alpha_s)=(0.97265,2.39\times10^{-3},-3.55\times10^{-4})$.
The posterior covariance has a direct dynamical interpretation. Variation in \(N_*\) and \(d\) moves points mainly along the pole attractor direction. Variation in \(c_2\) moves points across it by changing the canonical curvature, while variation in \(c_8\) changes the endpoint and the map between \(N_*\) and \(s_*\). The pair $(c_1,N_*)$ therefore dominates the width in $r$, whereas $c_2$ dominates the width in $n_s$. The wall coefficient remains weakly correlated with the CMB observables because its contribution is $1.53\times10^{-6}$ at the benchmark pivot. The covariance therefore separates topology, finite distance curvature, and exit dynamics into distinct directions in the ACT posterior distribution over coefficient space.

\inlinewidefigure{0.88}{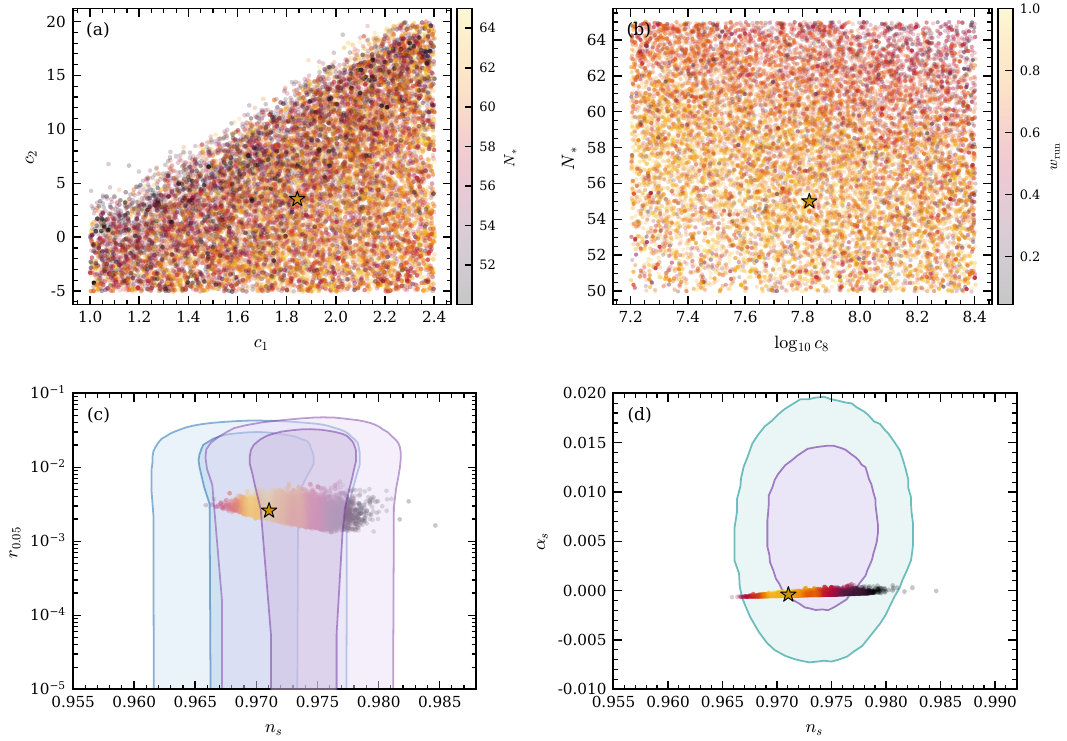}{Reduced branch distributions inferred from the ACT posterior products from 32,400 retained MCMC samples. The upper panels display the sampled coefficient directions, $N_*$, and the correlated scalar running weight $w_{\rm run}$. The lower panels locate the ensemble in the tensor and running planes. The star marks the benchmark. Sampling weights are reconstructed from the official ACT DR6.02 posterior chains. The smooth shaded boundaries provide visual context.}{fig:mcmcposterior}

\inlinewidefigure{0.88}{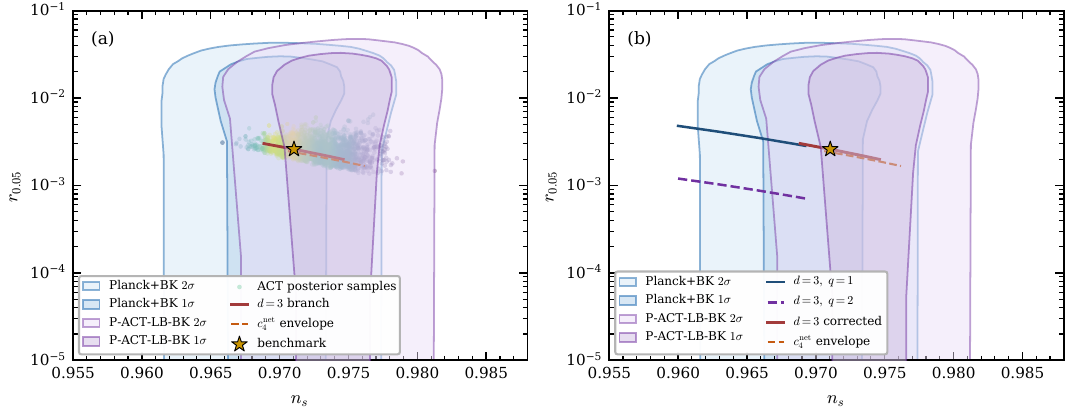}{Observable scan in the $(n_s,r)$ plane. The upper panel displays the ordering constrained \(d=3\) ACT posterior ensemble and the residual $c_4^{\rm net}$ envelope over the Planck, ACT, and BK18 visual boundaries. The lower panel compares the boundary attractor curves, the displaced $d=3$ trajectory, and the analytic $d=1$ reference branch.}{fig:nsr}

Figure~\ref{fig:futureforecast} makes the experimental reach explicit. The benchmark and the $50\le N_*\le65$ band of the $d=3$ branch lie in the small $r$ region probed by upcoming B mode searches. The nominal LiteBIRD and CMB-S4 sensitivities intersect the predicted band, placing the branch tensor scale within the design reach of the next generation.

\inlinewidefigure{0.86}{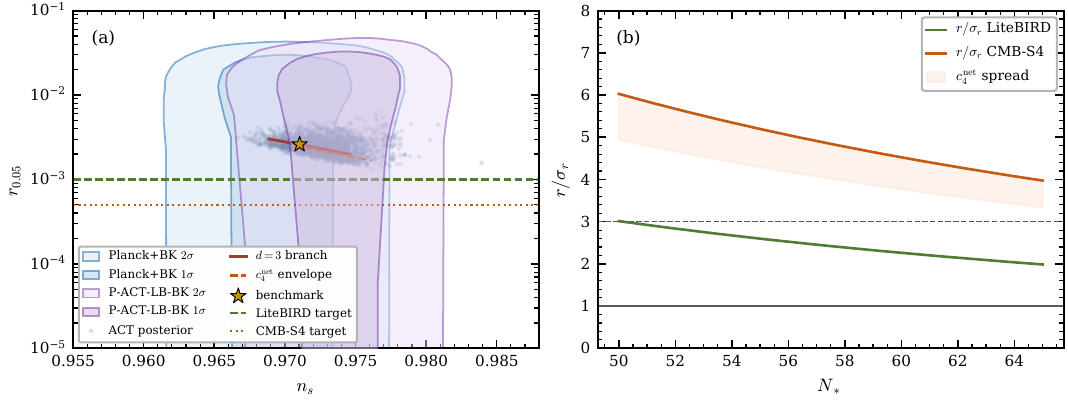}{Projected \((n_s,r)\) test. The current posterior boundaries are overlaid with the \(d=3\) ACT posterior ensemble, benchmark branch, residual $c_4^{\rm net}$ envelope, \(50\le N_*\le65\) band, and the quoted LiteBIRD and CMB-S4 tensor sensitivity targets.}{fig:futureforecast}

The running remains small and negative. Figure~\ref{fig:running} overlays the $d=3$ scan and residual $c_4^{\rm net}$ envelope with the official posterior $(n_s,\alpha_s)$ boundaries. The upward tilt shift is driven by Eq.~\eqref{eq:etashift}, while $\alpha_s$ remains of order $10^{-4}$.

\inlinewidefigure{0.92}{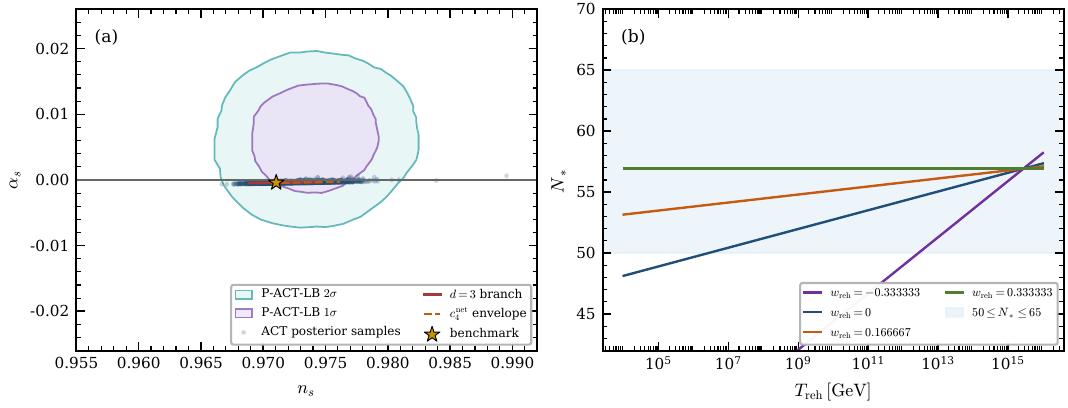}{Running and reheating tests. The upper panel displays the $d=3$ ACT posterior ensemble and the residual $c_4^{\rm net}$ envelope in the $(n_s,\alpha_s)$ plane with the ACT running boundaries. The lower panel displays the reheating temperature map from Eq.~\eqref{eq:reheat}.}{fig:running}

The number of e folds is related to a constant equation of state reheating phase through the standard matching relation~\cite{LiddleLyth:2000,Cook:2015}.
\begin{equation}
N_*\simeq57-{\frac{1}{4}}\log\left({\frac{V_*}{\rho_{\rm end}}}\right)
+{\frac{1-3w_{\rm reh}}{12(1+w_{\rm reh})}}\log\left({\frac{\rho_{\rm reh}}{\rho_{\rm end}}}\right)+\cdots .
\label{eq:reheat}
\end{equation}
We report the branch for $50\le N_*\le65$, spanning the conventional reheating interval from efficient radiation production to prolonged matter dominated reheating~\cite{Cook:2015,MartinRingeval:2014}. For the benchmark energy scale, $V_*=8.08\times10^{-11}$ and $\rho_{\rm end}=6.01\times10^{-11}$ in reduced Planck units. With $g_*=106.75$, the displayed reheating map spans $10^4\le T_{\rm reh}/{\rm GeV}\le2.8\times10^{15}$, the upper limit being the instantaneous reheating bound implied by \(\rho_{\rm end}\), and $w_{\rm reh}=0,1/6,1/4$. Matter dominated reheating yields $N_*=48.15,52.75,57.36$ at $T_{\rm reh}=10^4,10^{10},10^{16}\,{\rm GeV}$, respectively, while $w_{\rm reh}=1/4$ maps the temperature range near $55.18\le N_*\le57.02$. Across $50\le N_*\le65$, the benchmark trajectory moves from $(n_s,r)=(0.96887,3.01\times10^{-3})$ to $(0.97462,1.99\times10^{-3})$, so the low tensor branch retains the reheating dependence while preserving the branch ordering. The local curvature coefficient $c_2$ produces the ACT supported tilt shift inside the reheating interval.

The rolling saxion is a complex structure modulus, so the saxion postinflationary decay is a basic cosmological control check~\cite{Coughlan:1983,DeCarlos:1993,Banks:1993en}. Taking the oscillation mass at the inflationary scale, $m_s\sim H_*\simeq1.3\times10^{13}\,{\rm GeV}$, a gravitational decay width $\Gamma_s\simeq\kappa m_s^3/\mpl^2$ with $\kappa\sim(4\pi)^{-1}$ implies $T_{\rm dec}\simeq3\times10^9\,{\rm GeV}$. The inferred decay temperature lies far above the BBN scale and inside the reheating window displayed in Figure~\ref{fig:running}. The estimate ties the modulus decay to the high scale branch hierarchy used in the CMB calculation. A specified brane sector maps the total width into visible and axionic decay channels while leaving the inflationary observables unchanged.

The leading omitted terms in the period vector are proportional to $\ee^{2\pi it}$. The instanton contribution to the reduced potential satisfies the parametric bound
\begin{equation}
\left|{\frac{\Delta U_{\rm inst}}{U}}\right|\lesssim A_{\rm inst}\ee^{-2\pi s},
\label{eq:instantonbound}
\end{equation}
The coefficient \(A_{\rm inst}\) collects the period and flux prefactors. Appendix~C uses the conservative normalization $A_{\rm inst}=1$ to display the largest correction allowed by the exponential factor alone. At the pivot,
\begin{equation}
\log_{10}\left|{\frac{\Delta U_{\rm inst}}{U}}\right|\lesssim-222.10.
\end{equation}
The bound propagates to the scalar spectrum at leading order. Across the displayed CMB interval the corresponding fractional modulation remains below plotting precision, so the observable penumbral deformation is polynomial and is governed by the retained $s^{-2}$ term. Appendix~C displays the operator hierarchy and the instanton upper bound.

\section{Towers and multifield stability}

The de Sitter and distance conjectures formulate proposed quantum gravity restrictions on positive potentials and large field distances~\cite{Brennan:2017,Palti:2019}. To evaluate the de Sitter criteria on the solved branch, define the dimensionless gradient and Hessian ratios~\cite{Obied:2018,Ooguri:2019,Andriot:2018}
\begin{equation}
c_V={\frac{|\nabla V|}{V}},
\qquad
c_H=-\lambda_{\rm min}\left({\frac{G^{IK}\nabla_K\nabla_JV}{V}}\right),
\label{eq:dsdiag}
\end{equation}
where $\lambda_{\rm min}$ is the smallest eigenvalue of the covariant Hessian with one index raised by the field space metric. On the relaxed, negligibly turning branch, the normal eigenvalue is positive and large, so the minimum eigenvalue lies along the slow roll tangent and
\begin{equation}
c_V=\sqrt{2\epsilon}=\sqrt{\frac{r}{8}},
\qquad
c_H=-{\frac{V_{,TT}}{V}}=-\eta.
\end{equation}
At the $d=3$ pivot,
\begin{equation}
c_V=1.81\times10^{-2},
\qquad
c_H=1.40\times10^{-2}.
\end{equation}
Both ratios are below unity. The observable branch therefore does not satisfy an order one gradient criterion or a tachyonic Hessian criterion, as occurs generically for slowly rolling models with $\epsilon_V\ll1$ and $|\eta_V|\ll1$~\cite{AndriotRoupec:2018,RoupecWrase:2018}. Equation~\eqref{eq:dsdiag} evaluates these conjectural ratios directly. Transverse masses and compactification thresholds provide the independent effective field theory tests. The distance conjecture predicts an exponentially descending tower along an infinite distance trajectory~\cite{Ooguri:2007,Palti:2019}. Charge orbit constructions identify the states that realize this scaling near Calabi Yau boundaries~\cite{GrimmPaltiValenzuela:2018,GrimmLiPalti:2019}.

The EFT scale ordering is normalized using the stabilized volume. The branch mass and slow roll quantities are evaluated on the numerical background defined by Eq.~\eqref{eq:branchd3}, the string, KK and tower entries use the volume and boundary distance scalings stated below. In the four dimensional Einstein frame normalization used for the numerical hierarchy, the string and KK estimates are
\begin{equation}
{\frac{M_s}{\mpl}}\simeq {\frac{g_s^{1/4}}{\sqrt{4\pi\cV}}},
\qquad
{\frac{M_{\rm KK}}{\mpl}}\simeq {\frac{g_s^{1/4}}{\sqrt{4\pi}\,\cV^{2/3}}}={\frac{M_s}{\cV^{1/6}}},
\label{eq:MsKK}
\end{equation}
up to shape dependent factors of unity~\cite{Grana:2006,DouglasKachru:2007}. The asymptotic tower estimate is parameterized by the canonical boundary distance~\cite{Ooguri:2007,GrimmPaltiValenzuela:2018,GrimmLiPalti:2019},
\begin{equation}
\Delta\varphi=\sqrt{\frac{d}{2}}\log {\frac{s}{s_0}},
\qquad
M_{\tw}(s)=M_{\tw,0}\exp(-\lambda_{\tw}\Delta\varphi).
\label{eq:tower}
\end{equation}
For the displayed scale comparison we take $\lambda_{\tw}=\sqrt{2/3}$ and normalize $M_{\tw}(s_*)=M_{\rm KK}$ at $N_*=55$. The resulting curve compares the exponential distance scaling with the independently calculated KK and string thresholds. We define the ultraviolet compactification threshold by
\begin{equation}
\Lambda_{\rm UV}=\min(M_{\tw},M_{\rm KK},M_s).
\end{equation}
Across the CMB interval,
\begin{align}
\min {\frac{M_{\tw}}{H}}&=2.88\times10^2,&
\min {\frac{M_{\rm KK}}{H}}&=2.68\times10^2,\nonumber\\
\min {\frac{M_s}{H}}&=8.49\times10^2,&
\min {\frac{\Lambda_{\rm UV}}{H}}&=2.68\times10^2.
\end{align}
The Kaluza Klein scale is the lower compactification threshold at large volume, and the string scale is larger by \(\cV^{1/6}=3.16\). The branch mass $m_X/H=\sqrt{35}=5.92$ is the lower threshold relevant to the single field reduction. Compactification scales bound the four dimensional description, whereas the branch mass bounds the adiabatic truncation. Table~\ref{tab:schur} quantifies the associated curvature correction. Both thresholds remain separated from $H$ across the CMB interval. Appendix~C collects the finite distance potential ratios.

Multifield inflation treats trajectory bending and entropy modes through the covariant tangent and normal basis~\cite{Gordon:2000,GrootNibbelink:2000}. Heavy field effective theories quantify the sound speed, Schur complement and adiabaticity corrections induced by the orthogonal sector~\cite{Achucarro:2010,Senatore:2010,ShiuXu:2011}. The covariant background satisfies
\begin{equation}
D_t\dot\phi^I+3H\dot\phi^I+G^{IJ}\nabla_JV=0,
\qquad
3H^2={\frac{1}{2}}G_{IJ}\dot\phi^I\dot\phi^J+V.
\end{equation}
The bending rate and entropy mass are defined in Eq.~\eqref{eq:msdef}, with the heavy field correction organized as in the covariant effective theory~\cite{Achucarro:2012}. Along a displaced branch obeying $\Delta>0$, Eqs.~\eqref{eq:harmless1} and \eqref{eq:harmless2} and $p\le2$, the projected Hessian is positive in the normal direction and the bending correction is suppressed by
\begin{equation}
{\frac{4\Omega^2}{M_{\rm eff}^2}}\ll1.
\end{equation}
The benchmark has $\min(m_{\rm iso}^2/H^2)=34.9997$ and $\max(\Omega/H)=1.66\times10^{-4}$ throughout the CMB interval. The mixing ratio satisfies
\begin{equation}
\max {\frac{4\Omega^2}{M_{\rm eff}^2}}=3.15\times10^{-9}\label{eq:turnratio}.
\end{equation}
The ratio in Eq.~\eqref{eq:turnratio} bounds the displacement of the adiabatic spectrum below the displayed precision. The K\"ahler response is controlled independently by the Schur complement in Table~\ref{tab:schur}. Figure~\ref{fig:tower} displays the compactification thresholds and the covariant multifield ratios.

\inlinewidefigure{0.92}{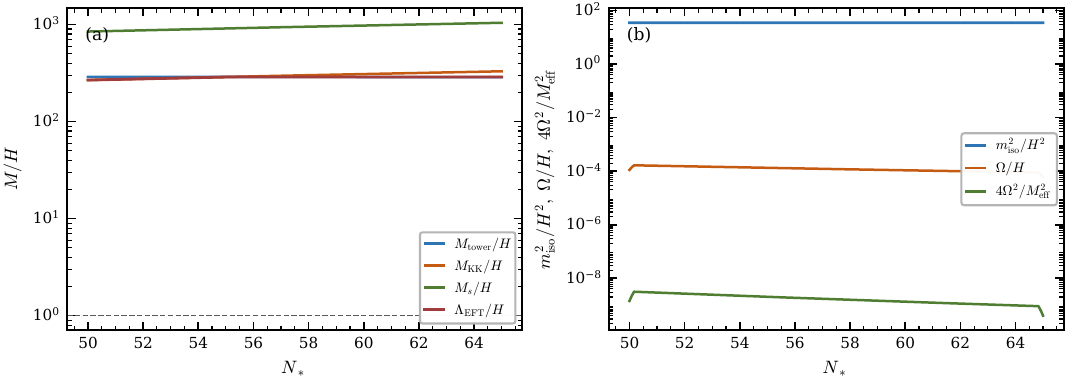}{Tower, cutoff, and multifield tests. The left panel displays the distance scaling estimate $M_{\rm tower}/H$, normalized at the pivot, together with $M_{\rm KK}/H$, $M_s/H$, and $\Lambda_{\rm EFT}/H$ along the displaced $d=3$ branch. The right panel displays $m_{\rm iso}^2/H^2$, $\Omega/H$, and $4\Omega^2/M_{\rm eff}^2$.}{fig:tower}

\section{Compact data and arithmetic conditions}

The active complex structure branch is the Greene Plesser mirror quintic branch described by Eq.~\eqref{eq:quinticpoly}~\cite{GreenePlesser:1990,Candelas:1991}. In a compact type IIB orientifold, a holomorphic involution $\sigma$ acts on the holomorphic three form, splits the third cohomology into even and odd subspaces, and restricts the three form fluxes to the odd lattice~\cite{GKP:2002,Grana:2006}
\begin{equation}
(f,h)\in H^3_-(Y_3,\mathbb Z)\equiv\Lambda_{\rm odd}.
\label{eq:oddflux}
\end{equation}
The D3 tadpole condition is~\cite{GKP:2002,AshokDouglas:2004}
\begin{equation}
N_{\rm flux}+N_{D3}=L,
\qquad
N_{\rm flux}=f^T\Sigma h.
\label{eq:tadfull}
\end{equation}
The orientifold or F theory lift determines \(L\). In an F theory description $L=\chi(Y_4)/24$. In an O3 and O7 description, O planes and curvature corrected D7 sources determine the negative D3 charge that is balanced by flux and mobile D3 branes~\cite{GKP:2002,Giryavets:2004}. The active flux vector in Eq.~\eqref{eq:fluxchoice} has $N_{\rm flux}=0$, leaving the tadpole capacity available to the spectator sector and localized sources.

Flux quantization and tadpole cancellation impose arithmetic restrictions on compact vacuum searches~\cite{DenefDouglas:2004,AshokDouglas:2004,Sethi:2017}. Strong tadpole requirements further constrain spectator stabilization and its available charge budget~\cite{Bena:2018}. The bounded active branch data below have $N_{\rm flux}=0$. A specified odd lattice realizes the active branch when the low charge, the coefficient image, and the spectator mass gap satisfy the simultaneous arithmetic system
\begin{align}
(f,h)&\in\Lambda_{\rm odd},
\qquad f^T\Sigma h+N_{D3}=L,
\label{eq:arith1}\\
\bm c_{\rm br}&=\mathcal C(f,h,S,T,\Pi,\sigma),
\label{eq:arith2}
\end{align}
\begin{equation}
\bm c_{\rm br}\equiv(c_1,c_2,c_4^{\rm net},c_8,A_p,B_{p+\nu}).
\end{equation}
The coefficient map $\mathcal C$ is computed by Eqs.~\eqref{eq:Aj} through \eqref{eq:c8match}. The orientifold involution selects the flux lattice and also determines the D3 source charge bound.  The penumbral branch theorem then tests the resulting coefficients for the stated compact data.

For the $d=3$ branch the working branch data are
\begin{align}
N_{\rm flux}&=0, \qquad d=3,
\qquad {\frac{\hat\xi}{2\cV}}=-1.074\times10^{-2},
\nonumber\\
(c_1,c_2,c_4^{\rm net})&=(1.8432,3.5257,0),
\nonumber\\
c_8&=6.6603\times10^7,
\nonumber\\
m_X^2/H^2&=35.
\label{eq:compactdata}
\end{align}
The geometric, flux, posterior, and stability calculations determine the entries in Eq.~\eqref{eq:compactdata}. The value \(d=3\) is the cubic Hodge degree of the maximally unipotent mirror quintic boundary~\cite{Candelas:1991,CattaniKaplanSchmid:1986}. The charge \(N_{\rm flux}=0\) follows from the bounded type IV weight conditions in Eq.~\eqref{eq:weightthm}. The representative in Eq.~\eqref{eq:fluxchoice} gives \(g_s=0.079861\) at the ACT inferred pivot. With \(\tau_0=100\), hence \(\mathcal V=10^3\), the BBHL ratio is \(\hat\xi/(2\mathcal V)=-1.074\times10^{-2}\), so the leading alpha prime correction is at the percent level~\cite{BBHL:2002}.

The ACT inferred reduced coefficients specify the full stationary branch. The coefficient \(c_1=1.8432\) measures the leading inverse saxion approach generated after no scale breaking. The ratio \(c_2/(c_1s_*)=2.35\times10^{-2}\) measures the finite distance curvature correction at the pivot. The endpoint coefficient \(c_8=6.6603\times10^7\) contributes \(1.53\times10^{-6}\) relative to the leading term at \(s_*\) and reaches \(0.949\) near \(s_{\rm end}=12.105\), producing the final steepening. For \(p=2\), \(d=3\), and \(m=1\), the choice \(m_X^2/H^2=35\) corresponds to \(A_2/V_0=8.75\). Together with \(X_v=-20/s\), the stationary equation gives \(B_3/A_2=40\). The normal field therefore has \(m_X/H=5.92\) and relaxes before the observable interval.

Equations~\eqref{eq:arith1} and \eqref{eq:arith2} state the orientifold arithmetic test for the ACT inferred reduced coefficients. The odd flux lattice must reproduce their coefficient image while satisfying the total tadpole and mass gap conditions. The ACT calculation determines the central reduced branch and its observational covariance. The arithmetic map then tests whether integral compactification data reproduce the coefficient vector.

\section{Conclusions}

The adiabatic direction is determined before the reduced potential is evaluated. In the reference axion valley, saxion response flattens the monodromic potential and the type IV gradient approaches $\gamma_{\rm LCS}=\sqrt6$~\cite{Lanza:2025}. The analytic equal weight model gives $\gamma_{\rm toy}=\sqrt2$. Here $\partial_XV=0$ eliminates $X=e-ma$ before the adiabatic projection. The saxion then lies along the tangent, while $X$ is the massive normal coordinate. Logarithmic saxion distance governs the canonical slope, and the benchmark reaches $\gamma_{\rm br}=1.80\times10^{-2}$. Equation~\eqref{eq:gammaratio} gives $\gamma_{\rm LCS}/\gamma_{\rm br}=2qN_*$ at leading order. The duration of the saxionic attractor therefore determines the slope suppression for the specified Hodge residue.

The stationary relation \(X_v(s)\sim s^{-\nu}\) links branch orientation, plateau energy, and entropy suppression. When \(p+2\nu>q\), the relaxation energy remains subleading to the leading plateau falloff. The coefficient \(A_p\) controls the heavy displacement, while the Hodge degree \(d\) determines the canonical pole residue. For the mirror quintic, the cubic Hodge norm gives \(d=3\) and \(\alpha_{\rm pole}=d/(3q^2)\). The tensor amplitude therefore retains the boundary class after saxion alignment. The coefficient \(c_2\) modifies the scalar curvature, whereas \(c_8\) determines the endpoint. Geometry, finite distance flux data, and exit dynamics consequently enter different derivatives of the reduced potential.

The integral period and flux calculation yields the polynomial coefficient map, the bounded active locus, and a weak coupling axiodilaton point with \(N_{\rm flux}=0\). The positive Hermitian Gram structure removes the growing type IV weights. The K\"ahler and uplift sectors then generate the Laurent coefficients of the plateau. ACT propagation at $N_*=55$ selects their central values before the observables are evaluated. With $\chi(Y)=+200$, the BBHL corrected volume sector gives $|\Delta\tau|/\tau_0=1.68\times10^{-3}$, $V_{\rm bar}/V_*=7.20$, and a Schur correction $|\Delta\eta|=1.66\times10^{-6}$. The normal mass, the volume eigenvalues, the compactification thresholds, and the bending rate retain the required hierarchy throughout the CMB interval.

At \(N_*=55\), the \(d=3\) branch yields
\begin{align}
n_s&=0.97105,
& r&=2.60\times10^{-3},\nonumber\\
\alpha_s&=-4.06\times10^{-4}.
\end{align}
The benchmark lies inside the $68.27\%$ highest density regions of the official ACT DR6.02 tensor and running posterior products~\cite{ACTDR6LCDM,ACTDR6Extended,ACTDR6Chains}. LiteBIRD and CMB-S4 probe the predicted tensor band~\cite{LiteBIRD:2022,CMBS4:2022}. Three inputs enter separate parts of the prediction. The type IV Hodge degree determines the canonical pole residue, stationary reduction determines the adiabatic direction, and the inferred Laurent coefficients determine the finite distance displacement of the scalar observables. Eqs.~\eqref{eq:gammaratio}, \eqref{eq:ACTbenchmarkcoeff}, and \eqref{eq:ACTbenchmarkobs} quantify these roles. Flux stationarity determines whether the logarithmic boundary metric lies along the inflaton tangent. When it does, the Hodge residue controls the leading tensor scale, while the integral coefficient map controls the departure from the boundary attractor. The $r\sim10^{-3}$ prediction therefore follows from the relation between the stationary tangent and the Hodge metric.

\begin{acknowledgments}
 
	TL is supported in part by the National Key Research and Development Program of China Grant No. 2020YFC2201504, by the Projects No. 11875062, No. 11947302, No. 12047503, and No. 12275333 supported by the National Natural Science Foundation of China, by the Key Research Program of the Chinese Academy of Sciences, Grant No. XDPB15, by the Scientific Instrument Developing Project of the Chinese Academy of Sciences, Grant No. YJKYYQ20190049, by the International Partnership Program of Chinese Academy of Sciences for Grand Challenges, Grant No. 112311KYSB20210012, and by the Henan Province Outstanding Foreign Scientist Studio Project, No.GZS2025008.

\end{acknowledgments}

\appendix

\section{Flux F term map}

The appendix records the algebra behind Eqs.~\eqref{eq:Aj} through \eqref{eq:Cell}. The goal is to make the coefficient extraction reproducible from the period vector, the symplectic flux vector and the axiodilaton value used in the main text. The nilpotent orbit expansion produces a polynomial period vector in the large complex structure coordinate $t=a+is$ up to exponentially suppressed instantons~\cite{Schmid:1973,Candelas:1991}. After choosing the symplectic basis used in Eq.~\eqref{eq:Pi}, the flux superpotential takes the polynomial form
\begin{align}
W(t,S)&=F(t)-S H(t),\\
F(t)&=\sum_{n=0}^{3}f_n^{\rm eff}t^n,
&
H(t)&=\sum_{n=0}^{3}h_n^{\rm eff}t^n .
\end{align}
The branch label determines $t_b=e/m+is$, and the monodromy invariant heavy coordinate is $X=e-ma$, so
\begin{equation}
t=t_b-{\frac{X}{m}},
\qquad
{\frac{\partial t}{\partial X}}=-{\frac{1}{m}} .
\end{equation}
For any holomorphic polynomial $P(t)$,
\begin{equation}
P(t_b-X/m)=\sum_{j\ge0}{\frac{(-1)^j}{j!m^j}}P^{(j)}(t_b)X^j .
\label{eq:appTaylor}
\end{equation}
The complex structure F term is
\begin{equation}
D_tW=\partial_tW+K_tW .
\end{equation}
Using $K_t=id/(2s)+\Order(s^{-2})$ at the boundary yields
\begin{equation}
D_tW(t_b-X/m,S)=\sum_{j\ge0}A_j(s)X^j,
\end{equation}
with
\begin{align}
A_j(s)=&\,{\frac{(-1)^j}{j!m^j}}
\partial_t^{j+1}W(t_b,S)
\nonumber\\
&+{\frac{(-1)^j}{j!m^j}}{\frac{id}{2s}}
\partial_t^jW(t_b,S)+\Order(s^{-2}).
\end{align}
The axiodilaton F term is
\begin{equation}
D_SW=-H(t)-{\frac{1}{S-\bar S}}W(t),
\end{equation}
so Eq.~\eqref{eq:appTaylor} yields
\begin{equation}
D_SW(t_b-X/m,S)=\sum_{j\ge0}B_j(s)X^j,
\end{equation}
with
\begin{equation}
B_j(s)={\frac{(-1)^j}{j!m^j}}\left[-H^{(j)}(t_b)-{\frac{1}{S-\bar S}}\partial_t^jW(t_b,S)\right].
\end{equation}
The no scale complex structure potential reads
\begin{equation}
V_{\rm cs}=e^{K_{\rm cs}+K_S}\left(K^{t\bar t}|D_tW|^2+K^{S\bar S}|D_SW|^2\right),
\end{equation}
The inverse metric obeys \(K^{t\bar t}=4s^2/d+\Order(s)\). Multiplying the two Taylor series yields
\begin{equation}
|D_tW|^2=\sum_{\ell\ge0}\left(\sum_{j+k=\ell}A_j(s)\overline{A_k(s)}\right)X^\ell,
\end{equation}
and the convolution applies to $D_SW$. Hence
\begin{align}
C_\ell(s)=e^{K_{\rm cs}+K_S}\Bigg[&{\frac{4s^2}{d}}
\sum_{j+k=\ell}A_j(s)\overline{A_k(s)}
\nonumber\\
&+K^{S\bar S}\sum_{j+k=\ell}B_j(s)\overline{B_k(s)}\Bigg]+\cdots .
\label{eq:CellAppendix}
\end{align}
The reduced coefficients entering the CMB branch follow by extremizing
\begin{equation}
V(X,s)=C_0(s)+C_1(s)X+C_2(s)X^2+\sum_{k\ge3}C_k(s)X^k .
\end{equation}
At leading order,
\begin{equation}
X_v(s)=-{\frac{C_1(s)}{2C_2(s)}}+\cdots,
\end{equation}
and
\begin{equation}
U(s)=C_0(s)-{\frac{C_1(s)^2}{4C_2(s)}}+\cdots .
\end{equation}
Matching to the stabilized branch in Eq.~\eqref{eq:branchd3} identifies the coefficients through
\begin{align}
c_1&=-{\frac{[s^{-1}]U}{V_0}}, &
c_2&={\frac{[s^{-2}]U}{V_0}},\nonumber\\
c_4^{\rm net}&=-{\frac{[s^{-4}]U}{V_0}}, &
c_8&=-{\frac{[s^{-8}]U}{V_0}} .
\end{align}
Equation~\eqref{eq:CellAppendix} maps integer flux data to the branch coefficients. In a compact orientifold the formula is evaluated on the odd flux lattice.

\section{BBHL and KKLT Hessian and barrier}

The K\"ahler sector starts from
\begin{equation}
K_K=-2\log Y,
\qquad
Y(\tau)=\tau^{3/2}+{\frac{\hat\xi}{2}},
\qquad
T=\tau+i\rho .
\end{equation}
The derivatives entering the one modulus supergravity potential are
\begin{align}
K_T&={\frac{1}{2}}{\frac{\partial K_K}{\partial\tau}}=-{\frac{3\sqrt{\tau}}{2Y}},\\
K_{T\bar T}&={\frac{1}{4}}{\frac{\partial^2 K_K}{\partial\tau^2}}
=-{\frac{3}{8\sqrt\tau Y}}+{\frac{9\tau}{8Y^2}},\\
K^{T\bar T}&=(K_{T\bar T})^{-1}.
\end{align}
With
\begin{equation}
W_K=W_0+A_{\rm np}\exp(-aT),
\end{equation}
the covariant derivative is
\begin{equation}
D_TW_K=-aA_{\rm np}e^{-aT}+K_T\left(W_0+A_{\rm np}e^{-aT}\right).
\end{equation}
On the real axis,
\begin{align}
|W_K|^2&=W_0^2+A_{\rm np}^2e^{-2a\tau}+2A_{\rm np}W_0e^{-a\tau}\cos(a\rho),\\
|D_TW_K|^2&={\cal A}(\tau)^2+{\cal B}(\tau)^2+2{\cal A}(\tau){\cal B}(\tau)\cos(a\rho),
\end{align}
where the radial functions are
\begin{equation}
{\cal A}(\tau)=K_TW_0,
\qquad
{\cal B}(\tau)=(-a+K_T)A_{\rm np}e^{-a\tau}.
\end{equation}
The uplifted potential used in the stabilization calculation is
\begin{equation}
V_K=e^{K_K}\left(K^{T\bar T}|D_TW_K|^2-3|W_K|^2\right)+{\frac{D}{(2\tau)^2}}.
\end{equation}
Since the $\rho$ dependence enters through $\cos(a\rho)$, the axis $\rho=0$ is invariant and the mixed derivative vanishes at the minimum.
\begin{equation}
\partial_\rho V_K|_{\rho=0}=0,
\qquad
\partial_\tau\partial_\rho V_K|_{\rho=0}=0 .
\end{equation}
The diagonal Hessian entries are
\begin{align}
\partial_\tau^2V_K&=\partial_\tau^2\left[e^{K_K}\left(K^{T\bar T}|D_TW_K|^2-3|W_K|^2\right)+{\frac{D}{(2\tau)^2}}\right],\\
\partial_\rho^2V_K&=-a^2e^{K_K}\left[2K^{T\bar T}{\cal A}{\cal B}-6A_{\rm np}W_0e^{-a\tau}\right]_{\tau=\tau_0} .
\end{align}
The real field kinetic term is $K_{T\bar T}[(\partial\tau)^2+(\partial\rho)^2]$, so the canonical eigenvalues are
\begin{equation}
m_\tau^2={\frac{\partial_\tau^2V_K}{2K_{T\bar T}}},
\qquad
m_\rho^2={\frac{\partial_\rho^2V_K}{2K_{T\bar T}}} .
\end{equation}
The barrier plotted in Figure~\ref{fig:kahlerbarrier} is obtained from
\begin{equation}
\partial_\tau V_K(\tau_{\rm max},0)=0,
\qquad
V_{\rm bar}=V_K(\tau_{\rm max},0)-V_K(\tau_0,0).
\end{equation}
The volume dressing of the branch energy follows from the factor $e^{K_K}=Y^{-2}$ in the supergravity potential.
\begin{equation}
V_{\rm br}(s,\tau)=V_0F(s){\frac{Y(\tau_0)^2}{Y(\tau)^2}}.
\end{equation}
Expanding the $\tau$ equation around $\tau_0$ yields
\begin{equation}
\Delta\tau(s)=-{\frac{\partial_\tau[V_0F(s)Y(\tau_0)^2/Y(\tau)^2]_{\tau_0}}{\partial_\tau^2V_K(\tau_0,0)}}+\Order\left({\frac{H^4}{m_\tau^4}}\right).
\end{equation}
The displayed Schur complement in Eq.~\eqref{eq:Schur} bounds the light field curvature shift from the overall volume subblock $(\tau,\rho)$. The branch normal direction $X$ is controlled by $m_X^2/H^2=35$, and the remaining flux sector directions enter through the positive mass condition in Eq.~\eqref{eq:cond3}.

\section{Background and observables}

The reduced branch is evolved in the canonically normalized variable
\begin{equation}
\varphi=\sqrt{\frac{d}{2}}\log s,
\qquad
{\frac{\dd\varphi}{\dd s}}=\sqrt{\frac{d}{2}}{\frac{1}{s}} .
\end{equation}
For any reduced potential $U(s)$,
\begin{align}
U_{,\varphi}&=\sqrt{\frac{2}{d}}sU_{,s},\\
U_{,\varphi\varphi}&={\frac{2}{d}}s(U_{,s}+sU_{,ss}),\\
U_{,\varphi\varphi\varphi}&=\sqrt{\frac{2}{d}}s{\frac{\dd}{\dd s}}\left[{\frac{2}{d}}s(U_{,s}+sU_{,ss})\right].
\end{align}
Together with the spectral definitions in Eq.~\eqref{eq:observabledefs}, these relations yield the slow roll quantities in Eqs.~\eqref{eq:epsr} through \eqref{eq:xis}. The two field branch obeys the standard covariant multifield background equations~\cite{Gordon:2000,GrootNibbelink:2000,Achucarro:2010}
\begin{equation}
D_t\dot\phi^I+3H\dot\phi^I+G^{IJ}\nabla_JV=0,
\qquad
3H^2={\frac{1}{2}}G_{IJ}\dot\phi^I\dot\phi^J+V .
\end{equation}
The field space trajectories in Figure~\ref{fig:fieldevol} are integrated with the covariant two field equations above. The invariant heavy coordinate obeys the linearized attractor equation
\begin{equation}
\delta X''+\left(3-\epsilon-2{\frac{s'}{s}}\right)\delta X'
+{\frac{m_X^2}{H^2}}\delta X=-J_v(s,s'),
\end{equation}
with $J_v=X_v''+(3-\epsilon-2s'/s)X_v'$. With $m_X^2/H^2$ above the values in Table~\ref{tab:stability}, the homogeneous decay controls the approach to the valley across the displayed initial conditions.

The polynomial and nonperturbative hierarchy $s^{-2}$ contribution remains visible throughout the CMB interval, whereas the exit wall becomes relevant after the CMB interval and the instanton remainder remains exponentially smaller.

Figure~\ref{fig:swamplandmap} records the finite distance ratios used to compare the solved branch with the distance conjecture scaling and the local potential derivatives.

\inlinewidefigure{0.92}{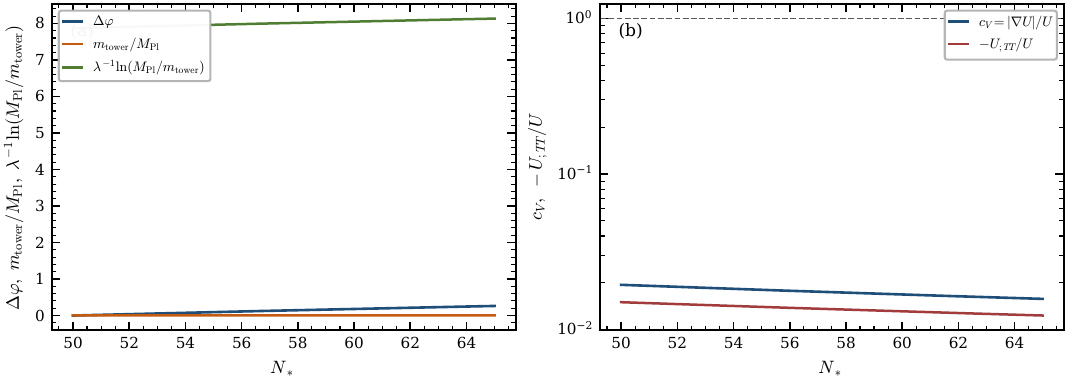}{Finite distance scale and potential ratios. The left panel displays $\Delta\varphi$, the tower estimate $m_{\rm tower}/M_{\rm Pl}$ normalized at the pivot, and $\lambda^{-1}\ln(M_{\rm Pl}/m_{\rm tower})$. The right panel displays $c_V$ and $-U_{;TT}/U$ across the CMB interval.}{fig:swamplandmap}

\section{Official ACT posterior construction}

The statistical calculation uses the public ACT DR6.02 posterior estimation archives~\cite{ACTDR6LCDM,ACTDR6Extended,ACTDR6Chains}. The weighted tensor density is reconstructed from the Planck and ACT lite chain with BK18, CMB lensing, and DESI BAO, while the weighted running density is reconstructed from the Planck and ACT lite NRUN chain after the prescribed burn in. Every model point is evaluated at $k_*=0.05\,{\rm Mpc}^{-1}$ from Eqs.~\eqref{eq:epsr} through \eqref{eq:xis}. Holding $N_*=55$, the tensor density defines the Metropolis target for $(c_1,c_2,\log_{10}c_8)$, and the NRUN density contributes the correlated weight because the products share CMB likelihoods. The componentwise marginal medians give Eq.~\eqref{eq:ACTbenchmarkcoeff}. Substitution into the branch equations gives Eq.~\eqref{eq:ACTbenchmarkobs}.

The smooth Planck, lensing, and BK18, Planck, ACT lensing, and BK18, and running boundaries are retained for visual comparison in Figures~\ref{fig:nsr}, \ref{fig:futureforecast}, \ref{fig:running}, and \ref{fig:mcmcposterior}. The sampler conditioned on $N_*$ uses the uniform prior $c_1\in[1.0,2.4]$, $c_2\in[-5,20]$, and $\log_{10}c_8\in[7.2,8.4]$, together with $|U_2/U_1|<0.25$, $|U_8/U_1|<5\times10^{-4}$, and $c_4^{\rm net}=0$. A second sampler includes $N_*\in[50,65]$ for reheating propagation. Each calculation retains 32,400 samples. Split $\hat R$ is evaluated for every sampled and derived coordinate.

\bibliographystyle{apsrev4-2-title}
\bibliography{v2refs}

\end{document}